\documentclass[11pt]{article}
\usepackage{amsmath}
\usepackage{amssymb}
\usepackage{amsthm}
\usepackage{multirow}
\usepackage{dsfont}
\usepackage{graphicx}
\usepackage{subcaption}
\usepackage{tabularray}
\usepackage{hyperref}
\usepackage{tikz}
\usepackage{blkarray,booktabs, bigstrut}
\usepackage{enumerate}
\usepackage[makeroom]{cancel}
\usepackage[utf8]{inputenc}
\usepackage{xcolor}
\usepackage[nottoc,numbib]{tocbibind}

\usepackage[
    inner=2.3cm,
    outer=2.3cm,
    top=2.7cm,
    marginparwidth=2.5cm,
    marginparsep=0.1cm
    ]
    {geometry}

\numberwithin{equation}{section}

\newcommand{\s}[1]{\mathcal{#1}}
\newcommand{\0}{\mathbf{0}}
\newcommand{\1}{\mathds{1}}

\newcommand{\norm}[1]{\left\lVert#1\right\rVert}
\newcommand{\sgn}{\text{sgn}}
\newcommand{\sign}{\text{sign}}
\newcommand{\A}{\mathcal{A}}
\newcommand{\V}{\mathcal{V}}
\newcommand{\M}{\mathcal{M}}
\newcommand{\G}{\mathcal{G}}
\newcommand{\E}{\mathcal{E}}
\newcommand{\T}{\mathcal{T}}
\newcommand{\F}{\mathcal{F}}
\renewcommand{\L}{\mathcal{L}}
\newcommand{\Li}[1]{\mathcal{L}^{(#1)}}
\renewcommand{\P}{\mathds{P}}
\newcommand{\R}{\mathds{R}}
\newcommand{\Rea}{\mathcal{R}}

\newcommand*{\QED}{\begin{flushright} $\square$ \end{flushright}}

\newtheorem{theorem}{Theorem}[section]
\newtheorem{proposition}[theorem]{Proposition}
\newtheorem{lemma}[theorem]{Lemma}
\newtheorem{corollary}[theorem]{Corollary}
\newtheorem{definition}[theorem]{Definition}

\theoremstyle{definition}
\newtheorem{remark}[theorem]{Remark}
\theoremstyle{definition}
\newtheorem{example}{Example}
\theoremstyle{definition}

\title{Sensitivity of steady states in networks with application to Markov chains and chemical reaction networks}
\author{Robin Chemnitz}
\date{December 2022}

\begin{document}

\maketitle

\begin{abstract}
    We consider steady states of dynamics that have an underlying network structure. We study how a steady state responds to small perturbations in the network parameters and how this sensitivity is connected to the network structure. We introduce a prototypical linear response equation and determine its sensitivity. This abstract result is applied to study the sensitivity of steady states in two common dynamics on networks: continuous-time Markov chains and deterministically modelled chemical reaction networks. For continuous-time Markov chains, we are able to efficiently compute the signs of the response in terms of the underlying network structure. The study of chemical reaction networks extends the sensitivity analysis to open systems with more complex network structures.
\end{abstract}

\section{Introduction}
Sensitivity analysis studies how an equilibrium of a dynamical system behaves under small perturbations in the system parameters. Many mathematical models whose sensitivity is studied in the applied sciences have underlying network structures, e.g.~food webs in ecological network analysis\cite{barzel2013universality, nakajima1992sensitivity, yodzis1988indeterminacy}, chemical reaction networks \cite{Feinberg, Brehm_Fiedler3, Nico_thesis, feliu2019sign} or cognitive radio networks \cite{zahmati2010steady} to name a few. Naturally, the underlying network structure greatly influence the sensitivity of an equilibrium by dictating how parameter perturbations propagate through the network. We study this connection in a universal setup, i.e.~determine which graph-theoretical structures influence the sensitivity of an equilibrium without specifying a concrete dynamics. We introduce a prototypical equilibrium equation with an underlying network structure and characterize its linear response. 

These general results are then applied to two types of dynamics with underlying network structures, namely continuous-time Markov chains and chemical reaction networks. While both of these are commonly modelled in a stochastic framework, they are easily formulated as systems of ordinary differential equations. These two types of dynamics already cover a wide range of mathematical models in application. Continuous-time Markov chains, in short CTMCs, are closed, i.e.~there is no in- or outflow of the system, and have a clear network structure associated to them. Chemical reaction networks can be open systems that admit in- and outflow. Additionally, the network structure is in general much more convoluted, as reactions may have multiple inputs and outputs that partially overlap.

The main results of the paper are the following theorems:
\begin{itemize}
    \item Theorem \ref{thm:general_main}: An explicit formula for the sensitivity of the prototypical equilibrium equation in terms of its underlying network structure;
    \item Corollary \ref{crl:signed_sens}: A complete description of the algebraic signs of the sensitivity of a CTMC, based on the underlying network structure, which can be computed efficiently;
    \item Theorem \ref{thm:sensitivity}: An explicit formula for the sensitivity at an equilibrium of a chemical reaction network in terms of a network structure derived from the set of reactions. This theorem demonstrates how the prototypical linear response can be applied to open systems with complex network structures.
\end{itemize}

In the sensitivity analysis of CTMCs we assume the existence of a unique stationary distribution and study how it responds to a small perturbation in one of the transition rates. This analysis can be carried out in two different ways. There is the numerical view-point, which answers the question of how large these responses can get, and there is the analytical view-point which analyses how the responses look like. The latter case, which is the one we will focus on, is one of the key steps in the control problem, i.e.~in order to achieve a desired response at the equilibrium one needs to understand how each possible perturbation of the system parameters influences the equilibrium. For discrete-time Markov chains, the numerical sensitivity of the stationary distribution has been studied by Schweitzer \cite{schweitzer_discrete_sens}, Hunter \cite{hunter1986stationary} and Funderlic and Meyer \cite{funderlic1986sensitivity}. More recently, Wang and Plechac \cite{wang2019steady} studied the sensitivity of an observable in a CTMC on a countable state space to the perturbation of a system parameter. 

In the last part of the paper, we apply the prototypical linear response to the sensitivity analysis of deterministically modelled chemical reaction networks. Chemical reaction networks mathematically model the evolution of chemicals components in a simplified model. These models may admit equilibria in which the concentrations of the components no longer fluctuate. Once again, the response of these equilibria to small perturbations in the system parameters is considered. We study the sensitivity in an algebraic way, like it has been done by Mochizuki and Fiedler \cite{Mochizuki_Fiedler1}, Feliu \cite{feliu2019sign} and Vassena \cite{Nico_thesis}. Our focus lies on the influence of the structure of the reaction network, i.e.~the way that the chemical components are linked to one another by the reactions. We introduce the \textit{influence graph} of a chemical reaction network that contains the structural information which is relevant for the sensitivity analysis and show that our results on the prototypical linear response generalize to this setting.

Section \ref{sec:prelim} introduces the necessary graph-theoretical definitions and theorems that will be needed for the following sections. Section \ref{sec:lin_response} presents the abstract sensitivity results of the prototypical linear response. Section \ref{sec:MC_Sens} applies these abstract results to CTMCs and derives a full characterization of the algebraic signs of the sensitivity of the stationary distribution. In Section \ref{sec:CRN} we briefly introduce chemical reaction networks and their sensitivity. We then show how the results of the prototypical linear response can be applied to the sensitivity of chemical reaction networks and present a description of the sensitivity in terms of the structure of the reaction network. We conclude with an outlook on open problems in Section \ref{sec:outlook}.

%. This means that the Markov chain must not be interpreted as a model of the system, but instead just admits similar behaviour. In this case, we construct a Markov chain that has the same sensitivity as the reaction network, even though their equilibria can be vastly different. This type of approach is possible since the sensitivity of an equilibrium of a differential equation only depends on its linearization in the equilibrium. Since CTMCs are easily described as a linear differential equation, they are a promising tool to study sensitivity. In Section \ref{sec:Outlook}, we will present some possible applications of this approach apart from chemical reaction networks.
\section{Graph theory preliminaries}\label{sec:prelim}
A directed graph $\G=(\V, \E)$ consists of a finite set of nodes  $\V$ and a set of edges $\E\subset \V\times \V$. We typically use letters $u,v,w$ for nodes while edges are denoted by the letter $j$. The direction of an edge is given by the order of the nodes, i.e.~$j=(uv)$ is an edge from $u$ to $v$ while $j'=(vu)$ is an edge from $v$ to $u$. In this paper, we will not allow self-loops, i.e.~edges of the form $(vv)$. A directed graph $\G=(\V, \E)$ is called \textit{strongly connected} if for each pair of nodes $u,v\in \V$ there is a path of directed edges in $\E$ that starts in $u$ and ends in $v$. A directed graph $\G=(\V, \E)$ is called \textit{weakly connected} if the graph is connected as an undirected graph i.e.~for each pair of nodes $u,v\in\V$ there is an undirected path of edges in $\E$ that starts in $u$ and ends in $v$ (the directions of the edges do not matter).\\

Trees are one of the most common structures of interest in graphs. A tree is a weakly connected subgraph with no undirected cycles. A tree that contains all nodes of the graph is called a spanning tree. A forest is a subgraph that consists of one or more vertex-disjoint trees. A forest is spanning if it contains all nodes of the graph. We will focus on a particular class of trees in directed graphs, so-called \textit{rooted trees}.
\begin{definition}
    Given a directed graph $\G = (\V, \E)$ and a root node $v_0\in \V$, we call a subgraph $\T = (\V', \E')$ a 'tree rooted in $v_0$' if $\T$ is a tree and every edge is directed towards $v_0$.
\end{definition}
For a more detailed description of the objects named above, we refer to \cite{fournier_graphs}. A rooted tree is also referred to as an \textit{arborescence} in the literature. Note that a rooted tree does not need to contain all vertices. In the case $\T$ contains all vertices of the graph, i.e.~$\V'=\V$, we call $\T$ a rooted spanning tree.
For any root node $v_0\in \V$ we denote the set of all spanning trees rooted in $v_0$ by 
\begin{equation*}
    \A^{v_0}:= \{ \T\subset \G, \: \T \text{ spanning tree rooted in }v_0\}.
\end{equation*}
The set of all spanning rooted trees in $\G$ is then defined as
\begin{equation*}
    \A:= \bigcup\limits_{v_0\in \V} \A^{v_0}.
\end{equation*}
If the graph is equipped with weights $(w_j)_{j\in \E}$ (in general, we may allow negative weights) we can define the weight of a rooted tree $\T$ (not necessarily spanning) by
\begin{equation*}
    \norm{\T} := \prod_{j\in \T} w_j.
\end{equation*}
Consequently, we define the quantities
\begin{align} \label{eq:total_foc}
    \begin{split}
        \norm{\A^{v_0}} &:= \sum_{\T\in \A^{v_0}}\: \norm{\T}, \\
        \norm{\A} &:= \sum_{v_0\in \V} \norm{\A^{v_0}}.
    \end{split}
\end{align}
In \cite{MCTT}, these quantities are referred to as the \textit{focus} of $v_0$, respectively the \textit{total focus} of the weighted graph $\G$. Note that these quantities naturally depend on the weights $w_j$ without being indicated in the notation. It should be clear from the context what the weights of the graph are such that the notation is not ambiguous.
We include a remark about the extension of rooted trees, which will be a useful tool in some of the proofs later. It comes in the form of the following lemma.
\begin{lemma}\label{lem:tree_extension}
    Let $\G=(\V, \E)$ be a directed graph and consider $\T'=(\V', \E')$ a tree rooted in some $v_0\in \V'$ with $\V'\neq \V$, i.e.~$\T'$ is not spanning. Under the assumption that $v_0$ is reachable from any $v\in \V$ by a directed path in $\G$, there is a spanning tree $\T$ which is rooted in $v_0$ such that $\T'$ is a subgraph of $\T$.
\end{lemma}

\textit{Proof:} We provide an explicit construction of $\T$. We start with $\T=\T'$. Consider any node $u\notin \T$. By assumption, there is a directed path $\gamma$ from $u$ to $v_0$. Starting at $u$, we add the edges of $\gamma$ to $\T$ until we arrive at a node that is already in $\T$ (at the latest, this happens at $v_0$). The resulting $\T$ is still a tree, since we could not have added a loop. Additionally, $\T$ is still rooted in $v_0$ since all edges that were added are directed towards $v_0$. This procedure can be repeated until all nodes are part of $\T$. By construction $\T$ is a spanning tree rooted in $v_0$ such that $\T'$ is a subgraph of $\T$. \QED
\begin{remark}\label{rem:A_non_empty}
    A special case of this Lemma states that in a strongly connected graph $\G$ there is at least one spanning tree rooted in $v_0$ for any $v_0\in \V$. If all edge-weights of the graph are positive, this implies that $\norm{\A^{v_0}}>0$ and thereby $\norm{\A}>0$.
\end{remark}

Finally, we define \textit{divided tree pairs}, also known in the literature as 2-trees \cite{2_trees}, which are substructures of directed graphs. These structures will be the central object in the results of the following sections.
\begin{definition}\label{def:j-dTp}
    Let $\G=(\V, \E)$ be a directed graph. For $w_1,w_2 \in \V$, we call $(\T_{w_1}, \T_{w_2})$ a 'divided tree pair', in short dTp, if 
    \begin{enumerate}[(i)]
        \item $\T_{w_i}\subset\G$ is a tree rooted in $w_i$, for $i=1,2$;
        \item $\forall\:u\in \V$ either $u\in \T_{w_1}$ (exclusive) or $u\in \T_{w_2}$.
    \end{enumerate}
    For an edge $j\in \E$, we call a dTp $(\T_{w_1}, \T_{w_2})$ a $j$-divided tree pair, in short $j$-dTp, if additionally
    \begin{enumerate}
        \item[(iii)] $\T_{w_1} \cup \T_{w_2} \cup j$ is weakly connected.
    \end{enumerate}
\end{definition}
There are a few observations to point out in this definition. The trees are allowed to consist of a single node. Condition $(ii)$ forbids $w_1=w_2$. Condition $(iii)$ implies that the edge $j$ has to connect the two trees and hence is part of neither of them. Additionally, for an edge $j=(uv)$ and any $j$-divided tree pair $(T_{w_1}, T_{w_2})$ we find that either $u\in T_{w_1}$, $v\in T_{w_2}$ or $v\in T_{w_1}$, $u\in T_{w_2}$. To distinguish these cases in formula, we use square brackets to interpret a Boolean term as a numerical value of $0$ or $1$. We write
\begin{equation}\label{eq:square_brackets}
    [u\in T_{w_1}] := \begin{cases} 1,\quad  u\in T_{w_1} \\ 0, \quad u \notin T_{w_1} \end{cases}.
\end{equation}

\section{Prototypical linear response}\label{sec:lin_response}
In this section we introduce a prototypical system of linear equations with an underlying network structure. Under mild assumptions, this system has a unique solution. We study how the solution responds to small perturbations in particular entries of the matrix governing the linear system, in particular in view of the underlying network structure.

Consider a directed graph $\G=(\V, \E)$ and a matrix $\L\in \R^{\V \times \V}$ with the properties
\begin{enumerate}[(i)]
    \item For nodes $u\neq v$ we have $\L_{vu} \neq 0 \iff (uv)\in \E$;
    \item $\L_{uu} = -\sum_{v \neq u} \L_{vu}$.
\end{enumerate}
The sparse structure of $\L$, i.e.~which entries are 0, uniquely characterizes the directed graph $\G$. Hence, we call $\G$ the underlying graph of $\L$. Additionally, $\L$ even provides weights for each edge. For $j=(uv)\in \E$ we set $w_j=\L_{vu}$. Whenever we refer to the weights of the graph $\G$, e.g.~by writing $\norm{A}$, we use the weights induced by $\L$. To emphasize this, we write $\G=(\V, \E, \L)$ for the weighted graph $\G$. Based on the strong similarity to the generator of a continuous-time Markov chain (see Section \ref{sec:MC_Sens}), we call $\L$ a \textit{generalized Laplacian}.

A generalized Laplacian $\L$ has the special property that each column sums to 0. In particular, this implies that the kernel of $\L$ is non-empty since the vector with 1 in ever entry is in the left-kernel of $\L$. We are interested in solutions $\mu \in \R^\V$ of the following system of linear equations which we call the \textit{equilibrium equation}
\begin{equation}\label{eq:proto_equation}
    \L \mu = 0, \quad \sum_{v\in \V} \mu_v = 1.
\end{equation}
In general, neither existence nor uniqueness of solutions of this system can be guaranteed. We introduce the matrix $\Li{v}\in \R^{\V \times \V}$ which is the matrix $\L$ with its $v$-row replaced by a row of $1$'s. Then any solution to (\ref{eq:proto_equation}) is also a solution to
\begin{equation}\label{eq:proto_Li}
        \Li{v} \mu = e_v,
\end{equation}
for any $v \in \V$. On the other hand, a simple calculation verifies that a solution $\mu$ of (\ref{eq:proto_Li}) is also a solution to (\ref{eq:proto_equation}). We compute the product of $\mu$ with the $v$-row of $\L$.
\begin{align*}
    \L_v \cdot \mu &= -\sum_{w\neq v} \L_w \cdot \mu \\
    &= 0.
\end{align*}
Hence, a vector $\mu \in \R^\V$ is a solution to (\ref{eq:proto_equation}) if and and only if it is a solution to the matrix equation (\ref{eq:proto_Li}) for any choice of $v\in \V$. Hence, we may refer to both of them as the equilibrium equation. The following proposition provides the condition under which there is a unique solution $\mu$ to the equilibrium equation.
\begin{proposition}\label{prop:det_Li}
    For any $v\in \V$ we find
    \begin{equation*}
        \det(\Li{v}) = (-1)^{|\V|-1} \norm{\A}.
    \end{equation*}
    Assuming that $\norm{\A}\neq 0$, there is a unique solution $\mu\in \R^\V$ to the equilibrium equation (\ref{eq:proto_equation}).
\end{proposition}

We proceed to study the sensitivity of the unique solution, i.e.~how it changes when perturbing a particular entry of $\L$. To guarantee existence and uniqueness of a solution $\mu\in \R^\V$, we assume $\norm{\A}\neq 0$. We may study the response to perturbation of arbitrary non-diagonal entries of $\L$, not only those of which correspond to an edge in $\E$. Let $\ell_{v^*u^*}$, for $u^*\neq v^*$, be the entry want to perturb, i.e.~replace $\ell_{v^*u^*}$ by $\ell_{v^*u^*} + \epsilon$. We do not require that $(u^*v^*)\in \E$. It is important to notice that increasing $\ell_{v^*u^*}$ by $\epsilon$ requires us to decrease the diagonal entry $\ell_{u^*u^*}$ by $\epsilon$ to preserve the properties of a generalized Laplacian. We define the perturbed Laplacian as
\begin{equation}\label{eq:L_Epsilon}
    \L(\epsilon) := \L + \epsilon \: \big( E_{v^*u^*} - E_{u^*u^*} \big),
\end{equation}
where $E_{vu}$ is the matrix with only $0$'s and one $1$ in row $v$ and column $u$.
For sufficienly small $\epsilon$, we still find $\norm{\A(\epsilon)}\neq 0$ and we may define $\mu(\epsilon)$ as the unique solution to
\begin{equation}\label{eq:mu_epsilon_def}
    \Li{v}(\epsilon) \mu(\epsilon) = e_v.
\end{equation}
It should be noted that $\Li{v}(\epsilon)$ is the matrix obtained by first adding the $\epsilon$ terms and then replacing the $v$-row by $1$'s. This may overwrite one of the $\epsilon$-terms if $v\in \{u^*, v^*\}$. We write $j^*=(u^*v^*)$ even if $j^*\notin \E$.
The response vector $\delta^{j^*}\in \R^\V$ to a perturbation of $j^*$ is defined as
\begin{equation}\label{eq:response_general}
    \delta^{j^*} := \frac{\partial \mu(\epsilon)}{\partial \epsilon}\Bigr|_{\epsilon=0}.
\end{equation}
\begin{theorem}\label{thm:general_main}
    Consider an generalized Laplacian $\L$ and let $\G=(\V, \E, \L)$ be the underlying weighted graph. We assume $\norm{\A}\neq 0$. When perturbing $j^*=(u^*v^*)$, not necessarily in $\E$, the response of the unique solution of the equilibrium equation (\ref{eq:proto_equation}) in a node $u'\in \V$ is given by
    \begin{equation}\label{eq:main_general}
        \delta_{u'}^{j^*} = \mu_{u^*} \frac{1}{\norm{\A}} \sum_{w\neq u'} \sum_{\substack{(\T_{u'},\T_w) \\ j^*\text{-dTp}}} (-1)^{[u^* \in \T_{u'}]} \norm{\T_{u'}}\norm{\T_w}.
    \end{equation}
\end{theorem}

Theorem \ref{thm:general_main} does not provide an efficient way to compute the entries of $\delta^{j^*}$ yet, as determining $j^*$-divided tree pairs is a computationally expensive task. However, the formula of the theorem provides insight into the relation of the graph structure of $\G$ and the responses $\delta^{j^*}_{u'}$. Additionally, this formula will be the central tool in the study of the sensitivity in Markov chains and chemical reaction networks in Section \ref{sec:MC_Sens} and \ref{sec:CRN}.

%The quantities $\norm{A},\mu_{u^*}, \norm{\T_{u'}}$ and $\norm{\T_w}$ do not need to be positive and $\mu_{u^*}$ might even be zero. Hence, a description of the sign of the responses like Corollary \ref{crl:signed_sens} is not derivable anymore. Instead, one can verify that by inverting the sign of each transition rate, i.e.~considering $-\L$, the response vector equals $-\delta^{j^*}$. Hence, in the generalized setting, the algebraic signs of the entries may only be $0$ or $\pm$, thus no interesting characterization of the signs is possible anymore.

%It would be an interesting topic to study the algebraic sign of the responses in a \textit{signed Laplacian} where each transition rate is fixed to be either positive or negative. In that setting, a characterization of the algebraic sign of the entries as in Corollary \ref{crl:signed_sens} is achievable, even though the conditions might be much more complex.

\subsection{Proofs}\label{sec:proofs}
This section contains the proofs to Proposition \ref{prop:det_Li} and Theorem \ref{thm:general_main}. The central tool of the proofs is the 'All Minors Matrix Tree Theorem' \cite{Chaiken_AMMTT}. It provides a formula for the determinant of a generalized Laplacian when deleting rows/columns from the matrix.

Consider a generalized Laplacian $\L$ with underlying weighted graph $\G=(\V, \E,\L)$. Additionally we fix an ordering of the vertices, i.e.~a bijection $\sigma:\V \to \{1,\hdots, |\V|\}$. For two sets of vertices $U,W\subset\V$, not necessarily disjoint, we denote by $\L(W\: | \: U)$ the submatrix of $\L$ obtained from deleting the rows indexed by $W$ and the columns indexed by $U$. For $U=\{u_1,\hdots,u_n\}$ and $W=\{w_1,\hdots,w_n\}$\footnote{We will only ever consider sets with $|U|=|W|$.} we may also write $\L(u_1,\hdots,u_n\:|\: w_1 \hdots w_n)$. The All Minor Matrix Tree Theorem can be stated as follows
\begin{theorem}\label{thm:AMMTT}
    \cite{Chaiken_AMMTT} For $U,W\subset \V$ with $|U|=|W|$ we find
    \begin{equation*}
        \det \big(\L(U \: | \: W)\big) = (-1)^{|\V|-|U|}(-1)^{\sum_{u\in U} \sigma(u)+  \sum_{w\in W} \sigma(w)} \sum_\F \textup{sgn} (\pi_\F) \norm{\F},
    \end{equation*}
    where the sum is over all spanning forests $\F$ of $\G$ such that
    \begin{enumerate}[(i)]
        \item $\F$ contains exactly $|W| = |U|$ trees;
        \item Each tree in $\F$ contains exactly one node in $U$ and exactly one node in $W$;
        \item Each tree in $\F$ is rooted in its node of $W$.
    \end{enumerate}
    $\F$ defines a bijection $\pi_\F: U \to W$ such that $\pi_\F(u) = w$ if and only if $u$ is in the tree of $\F$ which is rooted in $w$. As a bijection between ordered sets, $\pi_\F$ is a permutation whose signature is denoted by $\textup{sgn}(\pi_\F)\in \{-1, 1\}$. The product of the edge weights in $\F$ is denoted by $\norm{\F}$.
\end{theorem}
We will only apply this theorem in the cases $|W| = |U| = 1$ and $|W| = |U|=2$. In the first case, the forests $\F$ are simply spanning trees rooted in the node of $W$. In the latter case, the forests are divided tree pairs. More precisely, for $U=\{u_1, u_2\}$ and $W = \{w_1, w_2\}$ a forest $\F$ as in the theorem is a $(u_1u_2)$-divided tree pair $(\T_{w_1}, \T_{w_2})$.\\

\textit{Proof of Proposition \ref{prop:det_Li}:} Let $\L$ be a generalized Laplacian and fix any $v\in \V$. We compute the determinant of $\Li{v}$ via the Laplace-expansion in the $v$-row (whose entries are only $1$'s)
\begin{equation*}
    \det(\Li{v}) = \sum_{w\in\V}(-1)^{\sigma(v)+\sigma(w)}\det(\L(v \: | \: w)).
\end{equation*}
Theorem \ref{thm:AMMTT} yields
\begin{equation*}
   \det(\L(v \: | \: w)) = (-1)^{|\V|-1}(-1)^{\sigma(v)+\sigma(w)}\sum_\F \sgn(\pi_\F) \norm{\F}.
\end{equation*}
The sum runs over all spanning trees rooted in $w$. The bijection $\pi_\F$ maps between singleton sets and thus trivially has $\sgn(\pi_\F)=+1$. We conclude
\begin{align*}
\begin{split}
     \det(\Li{v}) &= (-1)^{|\V|-1}\sum_{w\in\V}\norm{\A^w} \\
     &= (-1)^{|\V|-1} \norm{\A}.
\end{split}
\end{align*}
Now let us assume $\norm{A}\neq 0$. Then, the matrix $\Li{v}$ is invertible for any $v\in \V$ and $\mu = (\Li{v})^{-1}e_v$ is the unique solution to the equilibrium equation (\ref{eq:proto_Li}). \QED

\textit{Proof of Theorem \ref{thm:general_main}:} Let $\L$ be a generalized Laplacian with underlying weighted graph $\G=(\V,\E)$ such that $\norm{A}\neq 0$. Let $j^*=(u^*v^*)$ be the entry of $\L$ we are perturbing. The solution $\mu(\epsilon)$ of the perturbed Laplacian $\L(\epsilon)$ is the uniquely determined by
\begin{equation*}
    \Li{v}(\epsilon) \mu(\epsilon) = e_v,
\end{equation*}
for any $v\in \V$ (compare (\ref{eq:mu_epsilon_def})). We choose $v=u^*$, the source node of $j^*$. This choice simplifies the matrix $\Li{u^*}(\epsilon)$, since one of the two $\epsilon$ terms is replaced by a $1$ (compare (\ref{eq:L_Epsilon})). We find
\begin{equation*}
    \Li{u^*}(\epsilon) = \Li{u^*}+\epsilon E_{v^*u^*}.
\end{equation*}
Hence, all $\mu(\epsilon)$ satisfy $f(\epsilon, \mu(\epsilon))=0$, where
\begin{align*}
\begin{split}
     f:\R\times\R^\V &\longrightarrow \R^\V \\
    (\epsilon,x) &\longrightarrow \Li{u^*}x + \epsilon x_{u^*}e_{v^*} - e_{u^*}.
\end{split}
\end{align*}
To apply the implicit function theorem, we verify that
\begin{equation*}
    \frac{\partial f}{\partial x}(0, \mu) = \Li{u^*},
\end{equation*}
is invertible. This is given by Proposition \ref{prop:det_Li} and the assumption $\norm{\A} \neq 0$. By implicit differentiation we obtain
\begin{align}\label{eq:implicit_diff}
    \begin{split}
            \frac{\partial \mu(\epsilon)}{\partial \epsilon}\Bigr|_{\epsilon=0} &= -\left(\frac{\partial f}{\partial x}(0, \mu)\right)^{-1} \frac{\partial f}{\partial \epsilon}(0, \mu) \\
            &= -\big(\Li{u^*}\big)^{-1} (\mu_{u^*} e_{v^*}).
    \end{split}
\end{align}
The left-hand-side of the equation is exactly the response vector $\delta^{j^*}$ whose entries we aim to compute. We can rewrite (\ref{eq:implicit_diff}) as
\begin{equation*}
    \Li{u^*} \delta^{j^*} = -\mu_{u^*} e_{v^*}.
\end{equation*}
Now let us fix a target node $u'\in \V$ in which we compute the response $\delta^{j^*}_{u'}$. Using Cramer's rule
\begin{equation*}
    \delta^{j^*}_{u'} = \frac{\det\big(B(u')\big)}{\det\big(\Li{u^*}\big)},
\end{equation*}
where $B(u')$ is the matrix obtained from replacing the $u'$-column of $\Li{u^*}$ by $-\mu_{u^*} e_{v^*}$. The determinant of $\Li{u^*}$ in the denominator was already computed in Proposition \ref{prop:det_Li}. We obtain
\begin{equation*}
    \delta^{j^*}_{u'} = (-1)^{|\V|-1}\frac{1}{\norm{\A}}\det\big(B(u')\big).
\end{equation*}
The determinant of $B(u')$ can be described using the Laplace-expansion in the $u'$-column. Since the $u'$-column is $-\mu_{u^*}e_{v^*}$ by construction, there is only one non-zero entry. The Laplace-expansion reads
\begin{equation}\label{eq:det_A}
    \det(B(u')) = -\mu_{u^*} (-1)^{\sigma(u') + \sigma(v^*)} \det\left(\Li{u^*}(v^* \: | \: u')\right).
\end{equation}
However, $\Li{u^*}$ is not a matrix we may apply Theorem \ref{thm:AMMTT} to, as it is not a generalized Laplacian. Therefore, we need to use yet another Laplace-expansion, this time in the $u^*$-row (which only contains $1$'s)
\begin{equation*}
    \det\left(\Li{u^*}(v^* \: | \: u')\right) = \sum_{\substack{w\in \V \\ w \neq u'}} (-1)^{\sigma_{\vee v^*} (u^*) + \sigma_{\vee u'}(w)} \det\left( \L(u^*, v^* \: | \: u', w)\right).
\end{equation*}
The terms of the form $\sigma_{\vee a} (b)$ denote the order of $b$ in $\sigma$ after deleting $a$. The order of $b$ only changes if $a$ came before $b$, i.e.~$\sigma(a)<\sigma(b)$ in which case it is reduced by one. We can rewrite the term as
\begin{equation*}
    \sigma_{\vee a} (b) = \sigma(b) - [\sigma(a) < \sigma(b)],
\end{equation*}
using the Boolean square bracket notation analogous to (\ref{eq:square_brackets}). We summarize
\begin{equation}\label{eq:delta_raw}
    \delta^{j^*}_{u'} = \mu_{u^*} \frac{1}{\norm{\A}} \sum_{\substack{w\in \V \\ w \neq u'}} \sign(u^*, v^*, u', w) \:\det( \L(u^*, v^* \: | \: u', w)),
\end{equation}
where, for spatial reasons, we abbreviated the signs of the summands as
\begin{equation*}
    \sign(u^*, v^*, u', w) = (-1)(-1)^{|\V|-1} (-1)^{\sigma(u^*) + \sigma(v^*) +  \sigma(u') + \sigma(w)} (-1)^{[\sigma(v^*)<\sigma(u^*)] + [\sigma(u') < \sigma(w)]}.
\end{equation*}
Now we apply Theorem \ref{thm:AMMTT} to compute the determinant of $\L(u^*, v^* \: | \: u', w)$
\begin{equation}\label{eq:det_L_two_out}
    \det(\L(u^*, v^* \: | \: u', w)) = (-1)^{|\V|-2} (-1)^{\sigma(u^*) + \sigma(v^*) +  \sigma(u') + \sigma(w)} \sum_\F \sgn(\pi_\F) \norm{\F}.
\end{equation}
As mentioned before, the sum runs over divided tree pairs $(\T_{u'}, \T_w)$ such that $u^*$ is in one and $v^*$ is in the other tree. Hence, the $\F$ are exactly the $j^*$-dTp of $\G$. To compute the sign of the bijection $\pi_\F$, we consider the two cases individually.\\
\textbf{Case 1.} $u^*\in \T_{u'}$

The bijection $\pi_\F$ is given by 
\begin{equation*}
    \pi_\F(u^*) = u' \qquad \pi_\F(v^*)=w.
\end{equation*}
One can check that the sign of this permutation is given by
\begin{equation*}
    \sgn(\pi_\F) = (-1)(-1)^{[\sigma(v^*)<\sigma(u^*)] + [\sigma(u') < \sigma(w)]}.
\end{equation*}
\textbf{Case 2.} $u^*\notin \T_{u'}$

The bijection $\pi_\F$ is given by 
\begin{equation*}
    \pi_\F(u^*) = w \qquad \pi_\F(v^*)=u'.
\end{equation*}
One can check that the sign of this permutation is given by
\begin{equation*}
    \sgn(\pi_\F) = (-1)^{[\sigma(v^*)<\sigma(u^*)] + [\sigma(u') < \sigma(w)]}.
\end{equation*}
The two cases are summarized by
\begin{equation*}
    \sgn(\pi_\F) = (-1)^{[u^* \in \T_{u'}]} (-1)^{[\sigma(v^*)<\sigma(u^*)] + [\sigma(u') < \sigma(w)]}.
\end{equation*}
Inserting this computation into (\ref{eq:det_L_two_out}) and again using the abbreviation of $\sign(u^*, v^*, u', w)$, we arrive at
\begin{equation}\label{eq:det_L_raw}
    \det(\L(u^*, v^* \: | \: u', w)) = \sign(u^*, v^*, u', w) \sum_{\substack{(\T_{u'},\T_w) \\ j^*\text{-dTp}}} (-1)^{[u^*\in T_{u'}]} \norm{\T_{u'}}\norm{\T_w}.
\end{equation}
Inserting (\ref{eq:det_L_raw}) into (\ref{eq:delta_raw}), we get
\begin{equation*}
    \delta^{j^*}_{u'} = \mu_{u^*} \frac{1}{\norm{\A}} \sum_{\substack{w\in \V \\ w \neq u'}} \sum_{\substack{(\T_{u'},\T_w) \\ j^*\text{-dTp}}} (-1)^{[u^*\in T_{u'}]} \norm{\T_{u'}}\norm{\T_w}.
\end{equation*}
This completes the proof. \QED

\section{Continuous-time Markov chains}\label{sec:MC_Sens}
Continuous-time Markov chains, in short CTMCs, are one of the most basic types of stochastic processes. We will briefly describe a CTMC in stochastic terminology before switching to a purely deterministic characterisation. A CTMC on a finite state space can be described as a random walk $(X_t)_{t\in\R_+}$ on a directed graph $\G=(\V, \E)$. At each point in time, the position of the random walk is given by a node in $\V$, i.e.~$X_t\in \V$. The trajectory of $X$ is assumed to be \textit{càdlàg}. Hence, the process $X$ jumps between the discrete states at discrete points in time. 
%Therefore, a CTMC on a discrete state space like we consider it is often referred to as a \textit{Markov jump process}.
The directed edges $\E$ indicate which transitions between the states are possible with positive probability. To each edge $(uv)\in\E$ we associate a positive rate at which $X_t$ is transitioning to $v$ from node $u$. For an explicit formulation of continuous-time Markov chains, we refer to \cite{Stroock2014CTMC}. We denote the transition rate of an edge $j=(uv)\in \E$ by $\ell_{vu}$ or simply $\ell_j$. Hence, the rates $(\ell_j)_{j\in \E}$ can be understood as weights of the edges in the graph. For $(uv)\notin \E$ we set $\ell_{vu}=0$. We define the \textit{Laplacian} $\L\in \R^{\V \times \V}$ of the CTMC as
\begin{equation}\label{eq:Laplacian}
    \L_{vu} = \begin{cases} \ell_{vu}, &u\neq v \\ -\sum_{w \neq v} \ell_{wv}, &u=v. \end{cases}
\end{equation}
The diagonal entries $\L_{vv}$ contain the total outgoing transition rates from a node $v$. This matrix is oftentimes referred to as the \textit{generator} of the CTMC. Note that the Laplacian $\L$ is also a generalized Laplacian as introduced in the previous section.

We shall study the probability distribution $p(t)\in \R^\V$ of the process $(X_t)_{t\in \R_+}$, which is defined as
\begin{equation*}
    p(t)_v = \P \big(X_t = v\big), \quad \forall v\in \V.
\end{equation*}
The Laplacian provides a purely deterministic representations of the Markov chain as a linear differential equation. Given an initial distribution $p_0\in \R^{\V}$ of $X_0$, the probability vector $p(t)$ satisfies
\begin{align}\label{eq:linear_ODE}
\begin{split}
    \Dot{p}(t) &= \L p(t), \\
    p(0) &= p_0.
\end{split}    
\end{align}
%Now the notation $\ell_{vu}$ for the rate from $u$ to $v$ should have become clearer as well since it allows for the convenient matrix-vector multiplication. The unintuitive notation can be avoided by considering matrix multiplication from the left instead of from the right, which is often done when working with Markov chains. However, in section \ref{sec:CRN} we present results in parallel to the works of Mochizuki, Fiedler, Brehm and Vassena \cite{Mochizuki_Fiedler1, Fiedler_Mochizuki2, Brehm_Fiedler3, Nico_thesis} in which matrix multiplication is done from the right. 
We will usually only characterize a CTMC by its Laplacian $\L$. The underlying graph structure can be derived from the positive values in $\L$. We call $\G=(\V, \E)$ the \textit{underlying graph} of the CTMC. The graph $\G$ is weighted with edge weights given by $(\ell_j)_{j\in \E}$ and we write $\G=(\V, \E, \L)$. A CTMC whose underlying graph is strongly connected is called \textit{irreducible}.

A probability vector $\mu\in \R^\V$ which is in the right kernel of $\L$ is called a \textit{stationary distribution}, in short stat.~dist. A well-known result is that under suitable conditions a stat.~dist.~exists and is unique.
\begin{theorem}\label{thm:ex_uni_mu}
    Consider an irreducible CTMC with Laplacian $\L$. Then, a stat.~dist.~exists and is unique. In formula, there is a unique vector $\mu\in\R^\V$ such that
    \begin{equation*}
        \L \mu = 0, \quad \sum_{v\in \V} \mu_v = 1.
    \end{equation*}
    Additionally, $\mu_v>0$ for any $v\in \V$.
\end{theorem}
For a proof of the theorem, we refer to \cite[Chapter 3]{norris1998markov}, in which much more general results are stated. 
\begin{remark}\label{rem:stat_dist_equil}
    Comparing the statement of Theorem \ref{thm:ex_uni_mu} to the equilibrium equation (\ref{eq:proto_equation}) shows that the stat.~dist.~of a CTMC is uniquely defined as the solution to the equilibrium equation of its Laplacian $\L$.
\end{remark}

\subsection{Sensitivity}
In this section we study how the stat.~dist.~$\mu$ of an irreducible Markov chain changes when slightly perturbing the transition rates. For simplicity, we study the effect of perturbing only one selected transition rate at a time. Our aim is to characterize how the structure of the underlying graph dictates the response of the stat.~dist.~to such perturbations.\\

Consider an irreducible CTMC with Laplacian $\L$ and let $\G=(\V, \E)$ be the underlying graph. Let $j^*=(u^*v^*) \in \E$ with rate $\ell_{v^*u^*}>0$ be the transition we want to perturb, i.e.~replace $\ell_{v^*u^*}$ by $\ell_{v^*u^*} + \epsilon$. It is important to notice that increasing $\ell_{v^*u^*}$ by $\epsilon$ requires us to decrease the leaving rate $\ell_{u^*u^*}$ by $\epsilon$ to preserve the Laplacian property. We define the perturbed Laplacian
\begin{equation}
    \L(\epsilon) := \L + \epsilon \: \big( E_{v^*u^*} - E_{u^*u^*} \big),
\end{equation}
where $E_{vu}$ is the matrix with only $0$'s and one $1$ in row $v$ and column $u$. For sufficiently small $\epsilon$ (to be precise $|\epsilon|<\ell_{v^*u^*}$) the Laplacian $\L(\epsilon)$ corresponds to an irreducible CTMC, which has a unique stat.~dist.~$\mu(\epsilon)$. Naturally, the underlying graph structure does not change under these perturbations. We are interested in the quantity
\begin{equation}\label{eq:delta_def}
    \delta^{j^*} := \frac{\partial \mu(\epsilon)}{\partial \epsilon}\Bigr|_{\epsilon=0},
\end{equation}
which we call the \textit{response vector} to a perturbation of $j^*$.
Equivalently, we may understand $\mu$ as a function of the transition rates $(\ell_j)_{j\in \E}$ and define the response as $\frac{\partial \mu(\ell)}{\partial \ell_{j^*}}$. 
%This quantity has been studied for discrete-time Markov chains by Schweitzer \cite{schweitzer_discrete_sens}, but only in a rather numerical setting. 
Our aim is to derive a description of the response vector $\delta^{j^*}$ in which the transition rates $(\ell_j)_{j\in \E}$ are left as variables. Hence, the response vector is a function $\delta^{j^*}: \R_+^{\E} \to \R^\V$ which only depends on the underlying (unweighted) graph $\G=(\V, \E)$. In particular, we study the \textit{signed response}, i.e.~whether the entries of $\delta^{j^*}$ are always positive / always negative / always zero or may change signs. Formally we define the algebraic sign of an entry $\delta_{u'}^{j^*}$ as
\begin{equation}\label{eq:algebraic_sign}
    \sign(\delta_{u'}^{j^*}) = \begin{cases} +, &\forall \: \ell\in \R_+^\E: \delta_{u'}^{j^*}(\ell) > 0 \\ -, &\forall \: \ell\in \R_+^\E: \delta_{u'}^{j^*}(\ell) < 0 \\
    0, &\forall \:\ell\in \R_+^\E : \delta_{u'}^{j^*}(\ell) = 0 \\
    \pm,  &\text{else.}\end{cases}
\end{equation}

We stress again that the algebraic sign of the entries of $\delta^{j^*}$ only depends on the graph structure of $\G$. Remark \ref{rem:stat_dist_equil} stated that the stat.~dist.~$\mu$ is the unique solution to the equilibrium equation of the Laplacian $\L$. Hence, the following theorem is a special case of Theorem \ref{thm:general_main}.
\begin{theorem}\label{thm:main_thm}
    Consider an irreducible CTMC with Laplacian $\L$ and let $\G=(\V, \E, \L)$ be the underlying weighted graph. When perturbing $j^*=(u^*v^*) \in \E$, the response of the stat.~dist.~in a node $u'\in V$ as defined in (\ref{eq:delta_def}) is given by
    \begin{equation}\label{eq:main_thm}
        \delta_{u'}^{j^*} = \mu_{u^*} \frac{1}{\norm{\A}} \sum_{w\neq u'} \sum_{\substack{(\T_{u'},\T_w) \\ j^*\text{-dTp}}} (-1)^{[u^* \in \T_{u'}]} \norm{\T_{u'}}\norm{\T_w}.
    \end{equation}
\end{theorem}
Note that in this setting Remark \ref{rem:A_non_empty} and Theorem \ref{thm:ex_uni_mu} guarantee that the quantities $\mu_{u^*}, \norm{\A}, \norm{\T_{u'}}$ and $\norm{\T_w}$ are all positive. This can be used to derive the following corollary characterizing the algebraic signs of the response entries in terms of the underlying graph structure.
\begin{corollary}\label{crl:signed_sens}
    Consider an irreducible CTMC with Laplacian $\L$ and let $\G=(\V, \E)$ be the underlying graph. When perturbing $j^*=(u^*v^*) \in \E$, the sign of $\delta_{u'}^{j^*}$ for $u'\in \V$ is 
    \begin{enumerate}[(i)]
        \item $+$, if and only if any path from $u^*$ to $u'$ contains $v^*$.
        \item $-$, if and only if there is no path from $u^*$ to $u'$ containing $v^*$.
        \item $\pm$, if and only if there are paths $\gamma_1$, $\gamma_2$ from $u^*$ to $u'$ such that $\gamma_1\ni v^* \notin \gamma_2$.
    \end{enumerate}
    In particular, the sign of $\delta_{u^*}^{j^*}$ is always $-$, while the sign of $\delta_{v^*}^{j^*}$ is always $+$. An entry which has sign $0$ cannot occur.  
\end{corollary}
\begin{remark}
    The existence of paths as described in Corollary \ref{crl:signed_sens} can be checked algorithmically with a complexity of $\mathcal{O}(|\V|\cdot|\E|)$. Hence, the algebraic sign of the response to a perturbation can be computed efficiently.
\end{remark}

\textit{Proof of Corollary \ref{crl:signed_sens}:} We consider the formula of the response in $u'\in \V$ to the perturbation of an edge $j^*=(u^*v^*)\in \E$
\begin{equation*}
    \delta_{u'}^{j^*} = \mu_{u^*} \frac{1}{\norm{\A}} \sum_{w\neq u'} \sum_{\substack{(\T_{u'},\T_w) \\ j^*\text{-dTp}}} (-1)^{[u^* \in \T_{u'}]} \norm{\T_{u'}}\norm{\T_w}.
\end{equation*}
Since all the transition rates $(\ell_j)_{j\in\E}$ are positive, $\mu_{u^*}\frac{1}{\norm{\A}}$ are strictly positive and do not influence the sign of $\delta^{j^*}_{u'}$. The products $\norm{\T_{u'}}\norm{\T_w}$ are monomials of degree $|\V|-2$ of the positive transition rates. Note that no monomial can appear twice and each monomial has a coefficient of either $+1$ or $-1$. Hence, the sign of $\delta^{j^*}_{u'}$ is determined by the existence of monomials with positive, respectively negative coefficient. This corresponds to the existence of a $j^*$-divided tree pair such that $u^*\in T_w$, respectively $u^*\in T_{u'}$. Indeed, if both types of $j^*$-dTp exists, the sign of $\delta^{j^*}_{u'}$ is $\pm$ while if only one type exist, the sign is $+$, respectively $-$. This requires the existence of at least one $j^*$-divided tree pair. As we will see, this is always the case.

To prove the corollary, it suffices to prove the following claims.\\
\textbf{Claim 1.} There is a $w\in \V$ and a $j^*$-dTp $(\T_{u'}, \T_w)$ with $u^*\in \T_w$ if and only if there is a directed path $\gamma$ from $u^*$ to $u'$ that contains $v^*$.\\
\textbf{Claim 2.} There is a $w\in \V$ and a $j^*$-dTp $(\T_{u'}, \T_w)$ with $u^*\in \T_{u'}$ if and only if there is a directed path $\gamma$ from $u^*$ to $u'$ that does not contain $v^*$.\\

Proving Claim 1: Consider a $j^*$-dTp $(\T_{u'}, \T_w)$ with $u^*\in \T_w$. This implies $v^*\in \T_{u'}$, and hence there is a directed path from $v^*$ to $u'$. Adding $j^*$ to the beginning of the path yields a directed path $\gamma$ from $u^*$ to $u'$ that contains $v^*$.

Consider a directed path $\gamma$ from $u^*$ to $u'$ that contains $v^*$ (This rules out $u'=u^*$). This path can be interpreted as a tree $\T'$ rooted in $u'$. By the extension Lemma \ref{lem:tree_extension}, we can find a spanning tree $\T$ rooted in $u'$ that contains $\T'$. Deleting the outgoing edge of $u^*$ splits $\T$ into a tree rooted in $u'$ and one rooted in $u^*$. Hence, it becomes a $j^*$-dTp $(\T_{u'}, \T_{u^*})$ for which it clearly holds that $u^*\in \T_{u^*}$ (in this case we have $w=u^*$).\\

Proving Claim 2: Consider a $j^*$-dTp $(\T_{u'}, \T_w)$ with $u^*\in \T_{u'}$. Hence, there is a directed path $\gamma$ from $u^*$ to $u'$ in $\T_{u'}$. Since $v^*\in \T_w$, this path does not contain $v^*$.

Consider a directed path $\gamma$ from $u^*$ to $u'$ that does not contain $v^*$ (This rules out $u'=v^*$). This path can be interpreted as a tree $\T'$ rooted in $u'$. By the extension Lemma \ref{lem:tree_extension}, we can find a spanning tree $\T$ rooted in $u'$ that contains $\T'$. Deleting the outgoing edge of $v^*$ splits $\T$ into a tree rooted in $u'$ an one rooted in $v^*$. Hence, it becomes a $j^*$-dTp $(\T_{u'}, \T_{v^*})$ for which $u^*\in \T_{u'}$ since $\gamma$ is still part of $\T_{u'}$ (in this case we have $w=v^*$). \\

Since $\G$ is strongly connected, there is at least one directed path from $u^*$ to $u'$ and therefore at least one $j^*$-dTp exists. Note that Claim 1 covers the case $u'=v^*$ Claim 2 covers the case $u'=u^*$.\QED

We illustrate the application of this Corollary in a toy example.
\begin{example}\label{ex:CTMC}
    We consider a CTMC with six states that we label from $A$ to $F$. The transitions between the states that may occur with positive probability are depicted in Figure \ref{fig:example_CTMC}(a). The Laplacian of the CTMC is written as
    \begin{equation*}
        \L = \begin{pmatrix}
            -(\ell_{BA} + \ell_{EA}) & 0 & 0 & \ell_{AD} & 0 & 0 \\ \ell_{BA} & -(\ell_{CB} + \ell_{DB}) & \ell_{BC} & 0 & 0 & 0 \\
            0 & \ell_{CB} & -\ell_{BC} & 0 & 0 & 0 \\
            0 & \ell_{DB} & 0 & -(\ell_{AD}+\ell_{FD}) & \ell_{DE} & 0 \\
            \ell_{EA} & 0 & 0 & 0 & -\ell_{DE} & \ell_{EF} \\
            0 & 0 & 0 & \ell_{FD} & 0 & -\ell_{EF}
        \end{pmatrix}.
    \end{equation*}
    The underlying graph $\G=(\V, \E)$ of the CTMC is given by
    \begin{align*}
        \V &:= \{A, B, C, D, E, F\}\\
        \E &:= \{(AB), (AE), (BC), (BD), (CB), (DA), (DF), (ED), (FE)\}.
    \end{align*}
    From Figure \ref{fig:example_CTMC}(a) it is easy to see that $\G$ is strongly connected. Hence the CTMC is irreducible and admits a unique stat.~dist.~whose sensitivity we may study. Corollary \ref{crl:signed_sens} determines the algebraic sign of each response entry $\delta_{u'}^{j^*}$. We compute the signs for each $j^*\in \E$ and $u'\in \V$ and write them in a matrix $\delta \in \{+,-,\pm\}^{\V \times \E}$. This matrix is depicted in Figure \ref{fig:example_CTMC}(b).

    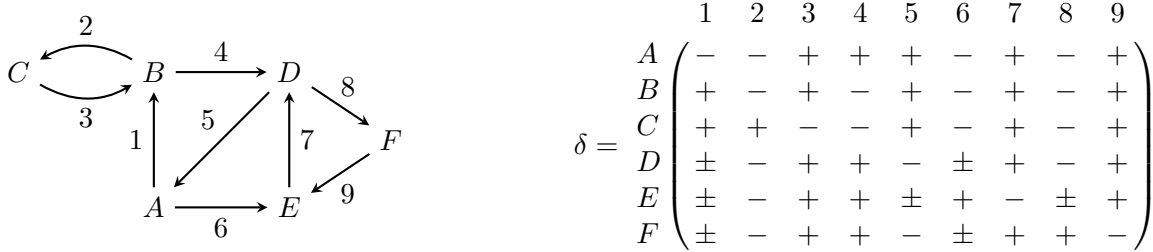
\begin{figure}[h]
    \centering
    \begin{subfigure}{0.42\textwidth}
    \vspace{6mm}
    \begin{tikzpicture}[node distance=1.8cm]
        \tikzstyle{arrow} = [thick,->,>=stealth]
        \node(C)[]{$C$};
        \node(B)[right of=C]{$B$};
        \node(A)[below of=B]{$A$};
        \node(D)[right of=B]{$D$};
        \node(E)[below of=D]{$E$};
        \node(F)[right of=E, xshift=-0.45cm, yshift=
        0.9cm]{$F$};
        
        \draw[arrow] (A) -- node [midway,left] {1} (B) ;
        \draw[arrow] (B)-- node [midway,above] {4} (D);
        \draw[arrow] (D) -- node[midway, above, xshift=-0.18cm]{5} (A) ;
        \draw[arrow] (E) -- node[midway, right]{7} (D);
        \draw[arrow] (D) -- node [midway,above, xshift=0.1cm] {8} (F) ;
        \draw[arrow] (F) -- node [midway, below, xshift=0.1cm]{9} (E);
        \draw[arrow] (A) -- node [midway, below]{6} (E);
        \draw[arrow] (B) to [out=150, in=30] node[midway, above]{2} (C);
        \draw[arrow] (C) to [out=-30, in=-150] node[midway, below]{3} (B);
    \end{tikzpicture}
    \vspace{6mm}
    \caption{Network structure of the CTMC.}
    \end{subfigure}%
    \begin{subfigure}{0.52\textwidth}
    \begin{equation*}
        \delta = \bordermatrix{
          & 1 & 2 & 3 & 4 & 5 & 6 & 7 & 8 & 9 \cr
        A & - & - & + & + & + & - & + & - & + \cr
        B & + & - & + & - & + & - & + & - & + \cr
        C & + & + & - & - & + & - & + & - & + \cr
        D &\pm& - & + & + & - &\pm& + & - & + \cr
        E &\pm& - & + & + &\pm& + & - &\pm& + \cr
        F &\pm& - & + & + & - &\pm& + & + & - \cr
        }
    \end{equation*}
    \caption{Algebraic signs of the response entries $\delta_{u'}^{j^*}.$}
    \end{subfigure}
    \caption{(a) Letters $A$ to $F$ denote the states of the CTMC. Possible transitions are labelled with numbers 1 to 9. (b) For $j^*\in \E$ and $u'\in \V$ the entry $\delta_{u'j^*}$ contains the algebraic sign of the response entry $\delta_{u'}^{j^*}$.}
    \label{fig:example_CTMC}
    \end{figure}
    Note that most of the signs of the response entries are determined, i.e.~either $+$ or $-$ while only a few entries are of undetermined sign $\pm$. A special response pattern arises for edges which are the only outgoing edge of a node, namely edges $3,7$ and $9$. Perturbation in these edges lead to a negative response in the node they are originating from and to a positive response in all other nodes. This is in accordance with Corollary \ref{crl:signed_sens}. \QED
\end{example}

\section{Chemical reaction networks}\label{sec:CRN}
In this section we present an application of the results on the prototypical linear response of generalized Laplacians to deterministically modelled chemical reaction networks. 
%For chemical reaction networks it is a common approach to model the system as a Markov chain population dynamic \cite{CRN_CTMC_Anderson}. However, this is not the type of application we are aiming for in this section. Instead, we construct a generalized Laplacian to imitate the linearization of a reaction network in a given equilibrium. As we will see, the sensitivity of the chemical reaction network can be derived from the sensitivity of the constructed Laplacian and computed using the results from the previous section.\\
In this brief introduction to chemical reaction networks, we closely follow the work of Vassena \cite{Nico_thesis}, which summarizes the framework of Mochizuki, Fiedler and Brehm \cite{Mochizuki_Fiedler1, Fiedler_Mochizuki2, Brehm_Fiedler3}. 

A reaction network consist of a set of reactants $\M$ and set of reactions $\Rea$ between them. A reaction $j\in\Rea$ might look like
\begin{equation*}
    j: \: A + 2 B \longrightarrow 2C + D,
\end{equation*}
where $A,B,C,D\in \M$ are reactants. We call the reactants on the left side of the reaction the \textit{inputs} and those on the right side the \textit{outputs}. A reactant can be both input and output of one and the same reaction. In the general form, a reaction $j$ can be represented as
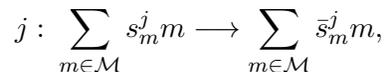
\begin{equation*}
    j: \: \sum_{m\in \M} s^j_m m \longrightarrow \sum_{m\in \M} \bar{s}^j_m m,
\end{equation*}
where $s^j_m,\:\bar{s}^j_m$ are non-negative values. We call them the \textit{stoichiometric coefficients}. Most of the times, the stoichiometric coefficients are integer-valued, however this is not required. In this general definition a reactant $m$ is an input if $s^j_m \neq 0$ and an output if $\bar{s}^j_m\neq 0$. Reactions that have at most one input and at most one output are called \textit{monomolecular}. If every reaction of a reaction network is monomolecular, we call it a monomolecular reaction network. There are two special types of reactions that we address explicitly. Those without inputs, which we call \textit{feed-reactions} and those without output, which we call \textit{exit-reactions}. We write these using the notation of the \textit{zero-complex} $\0$ \cite{Feinberg}, i.e.~$j:\0 \to \cdot$, respectively $j:\cdot \to \0$. Feed reactions can be understood as a constant inflow of certain reactants into the system, while exit reactions can indicate the production of a substance which is not part of the system, e.g. biomass or energy.

Writing $s^j$ and $\bar{s}^j$ as vectors in $\R^\M$ we can define the \textit{stoichiometric matrix} $S\in\R^{\M\times \Rea}$, whose $j$-column is given by
\begin{equation*}
    S^j = \bar{s}^j - s^j.
\end{equation*}
With this definition, we can describe the reaction network as a system of ordinary differential equations (ODEs). The quantities we consider are the concentrations $x_m$ of the individual reactants $m\in \M$. The values of the concentrations are not normalized, i.e.~they are not required to sum to 1 and may take arbitrary non-negative values. The temporal evolution of the concentration vector $x\in \R^\M$ is described by the potentially non-linear differential equation
\begin{equation}\label{eq:diff_equation}
    \Dot{x} = S r(x),
\end{equation}
where $r:\R^\M \to \R^\Rea$ are the reaction rates. We impose a few constraints on the reaction rates $r$. The rate $r_j(x)$ of a reaction $j$ is assumed to be differentiable, non-negative and only dependent on the input reactants of $j$. Additionally, the rate of $j$ increases monotonically with the concentration of the inputs. We can specify these constraint via the derivatives of $r$. Namely, we require that for any non-negative concentration vector $x\in \R_{\geq 0}^\M$
\begin{equation}\label{eq:rate_constraint}
    \frac{\partial r_j}{\partial x_m}(x) > 0 \quad \iff \quad m \text{ is an input of } j.
\end{equation}
In particular, this implies that feed reactions have a constant rate since their reaction rate does not depend on the concentration of the reactants. Common choices of the reaction rate function are defined by the mass-action kinetics or Michaelis-Menten kinetics, both of which satisfy the condition posed in (\ref{eq:rate_constraint}). However, we do not focus on any specific choice of the reaction rate function but aim to study the sensitivity of equilibria for arbitrary reaction rate functions.

An equilibrium of the system is a vector $x^*\in \R_{\geq 0}^\M$ for which
\begin{equation}\label{eq:equilibrium_cond}
    Sr(x^*) = 0.
\end{equation}
Existence and uniqueness of equlibria is a topic that can be studied on its own. Here, we assume the existence of an equilibrium $x^*$ and study its sensitivity as introduced by Mochizuki and Fiedler in \cite{Mochizuki_Fiedler1}. In particular, we describe the response of the equilibrium when the rate of a particular reaction $j^*$ is increased. In practice, this can be achieved by adding a catalyst of the reaction to the system. Mathematically, we add an $\epsilon$ to the reaction rate of $r_{j^*}$. We write $r_{j^*}(\epsilon, x) = r_{j^*}(x) + \epsilon$. All other reactions are assumed to be unaffected by $\epsilon$. Hence, we write the perturbed reaction rate vector by
\begin{equation*}
    r(\epsilon, x) := r(x) + \epsilon e_{j^*}.
\end{equation*}
We denote the equilibrium corresponding to $r(\epsilon, \cdot)$ by $x^*(\epsilon)$, i.e.
\begin{equation*}
    0 = Sr(\epsilon, x^*(\epsilon)).
\end{equation*}
In \cite{Mochizuki_Fiedler1} the response vector $\Delta^{j^*}\in \R^\M$ was introduced which is the response of the equilibrium concentrations to perturbation of the reaction $j^*$
\begin{equation*}
    \Delta^{j^*}:=\frac{\partial x^*(\epsilon)}{\partial \epsilon} \Bigr|_{\epsilon=0}
\end{equation*}
In \cite{Brehm_Fiedler3} it was shown that $x^*(\epsilon)$, and thereby the response $\Delta^{j^*}$, are well-defined under non-degeneracy of the Jacobian of (\ref{eq:diff_equation}) at the equilibrium $x^*$. To compute this Jacobian we define
\begin{equation}\label{eq:def_R}
    R_{jm} := \frac{\partial r_j}{\partial x_m} (x^*),
\end{equation}
the derivatives of the reaction rates at the equilibrium. Writing these values as a matrix $R\in \R^{\Rea \times \M}$, the Jacobian of the system is then given by $SR$. Hence, we require
\begin{equation}\label{eq:non_degeneracy}
    \det(SR) \neq 0.
\end{equation}
By implicit differentiation of the function
\begin{align*}
\begin{split}
    f:\R\times\R^\M &\longrightarrow \R^\M \\
    (\epsilon, x) &\longrightarrow Sr(\epsilon, x)
\end{split}
\end{align*}
in the point $(0, x^*)$, we get
\begin{align}\label{eq:delta_with_factor}
\begin{split}
    \frac{\partial x^*(\epsilon)}{\partial \epsilon} \Bigr|_{\epsilon=0} &= - \left( \frac{\partial f}{\partial x} (0, x^*) \right)^{-1} \left(\frac{\partial f}{\partial \epsilon} (0, x^*) \right) \\
    &= - (SR)^{-1} S e_{j^*}.
\end{split}
\end{align}
Using the previously introduced notation, we arrive at a formula for the response vector
\begin{equation}\label{eq:metabolite_response}
    \Delta^{j^*} = -(SR)^{-1}S^{j^*}.
\end{equation}
Note that the response vector does not depend on the actual values of the reaction rate $r(x^*)$, but only on the derivatives $R_{jm}$ at the equilibrium. Similar to the CTMC setting in which we considered arbitrary positive transition rates $(\ell_j)_{j\in\E}$, we leave the non-zero values $R_{jm}$ as positive parameters and aim to describe the response vector as a function of these values symbolically. By (\ref{eq:rate_constraint}), an entry $R_{jm}$ is a non-zero variable if and only if $m$ is an input to reaction $j$. The non-degeneracy condition (\ref{eq:non_degeneracy}) can be understood as the determinant being non-zero as a function of the values in $R_{jm}$. In \cite[Theorem 2.1]{Brehm_Fiedler3} an analysis of this condition is provided. In \cite{Mochizuki_Fiedler1, Brehm_Fiedler3} it is described under what condition $\Delta^{j^*} \neq 0$ holds \textit{algebraically}, i.e.~as a function. A function is considered algebraically non-zero if there is at least one valid input that yields a non-zero output. The work of Vassena \cite{Nico_thesis} then addressed the question of \textit{signed sensitivity}, i.e.~determining the algebraic sign (compare (\ref{eq:algebraic_sign})) of $\Delta^{j^*}$.

It should be mentioned that in a system with, for example, the mass-action kinetic, the derivatives $R_{jm}$ can not be chosen arbitrarily, since they are tied to the particular equilibrium $x^*$. This does not make any of the algebraic statements about $\Delta$ wrong, but makes them vastly over-generalized.

\iffalse
We make one more comment on \textit{reactant perturbations}. These perturbations describe the external input of a reactant $m$. However, this can be modelled easily with the existing framework we just introduced. A reactant perturbation simply corresponds to the perturbation of a feed-reaction $j^*:\0 \to m$. In that case $S^{j^*}=e_{m}$ and we obtain
\begin{equation}\label{eq:feed_perturbation}
    \Delta^{j^*} = - (SR)^{-1}e_{m}.
\end{equation}
Since both cases are covered by (\ref{eq:metabolite_response}), we do not consider reactant perturbations separately for reaction networks. It should be mentioned that the feed-reaction $j^*:\0\to m$ does not have to be a reaction in $\Rea$ in order to study its sensitivity. Formula (\ref{eq:feed_perturbation}) still applies even if $j^*\notin\Rea$.
\fi

\subsection{Sensitivity using generalized Laplacians}\label{sec:application}
As seen in the previous section, the sensitivity of a reaction network depends solely on the linearization $SR$ in a given equilibrium. In this section, we will interpret $SR$ as part of a generalized Laplacian $\L$ and obtain sensitivity results similar to Theorem \ref{thm:general_main}. 

Consider a reaction network with set of reactants $\M$ and set of reactions $\Rea$. Let $x^*$ be a given equilibrium of the differential equation (\ref{eq:diff_equation}) and let $SR$ be the linearization in $x^*$. We assume $\det\big(SR\big)\neq 0$. We construct a generalized Laplacian $\L$ on the set $\M\cup \{\0\}$ in the following way
\begin{equation} \label{eq:L_block}
\L = \left(\begin{array}{c|c c c}
    0& \multicolumn{3}{c}{-\1 SR} \\ \hline
    \multirow{3}{*}{0} & & & \\
    & & SR & \\ 
    & & & 
\end{array}\right),
\end{equation}
where the first row/column corresponds to the state $\0$. The symbol $\1$ is used to denote the vector of appropriate shape that consists only of $1's$. Hence, $\1SR$ is a row vector containing the column sums of $SR$ and (\ref{eq:L_block}) indeed defines a generalized Laplacian, i.e.~all columns sum to 0. We call $\G=(\M \cup \{\0\}, \E)$, the underlying graph of $\L$, the \textit{influence graph} of the reaction network in an equilibrium. Let us stress that the graph structure of $\G$ depends neither on the chosen equilibrium $x^*$ nor on the values $R_{jm}$, but purely on the set of reactants $\M$ and set of reactions $\Rea$. However, given an equilibrium $x^*$ and choice of values $R_{jm}$, we can equip $\G$ with weights using the generalized Laplacian constructed in (\ref{eq:L_block}). We write $\G=(\V, \E, \L)$ for the weighted influence graph. Before we continue, let us provide some intuition for the generalized Laplacian (\ref{eq:L_block}) and the influence graph, in particular which edges lie in the set $\E$. We consider an example reaction network.

\begin{example}\label{ex:CRN}
Consider the chemical reaction network defined by
\begin{align*}
\begin{split}
    \M &:= \{A,B,C\}\\
    \Rea &:= \{j_0:\0\to A,\: j_1:A\to B, \:j_2:A+B\to C,\: j_3:C\to \0\}.
\end{split}
\end{align*}
See Figure \ref{fig:example}(a) for a schematic drawing of the reaction network. For simplicity, we label the reactions $j_0,\hdots, j_3$ with their respective index $0,\hdots,3$.
\begin{figure}
    \centering
    \begin{subfigure}{0.46\textwidth}
    \centering
    \begin{tikzpicture}[node distance=1.2cm]
        \tikzstyle{arrow} = [thick,->,>=stealth]
        \node(0in)[]{$\0$};
        \node(A)[right of=0in]{$A$};
        \node(B)[below of=A]{$B$};
        \node(center)[right of=A, xshift=-0.3cm, yshift=-0.6cm]{};
        \node(2)[right of=A, xshift=-0.3cm, yshift=-0.2cm]{2};
        \node(C)[right of=center, xshift=-0.3cm]{$C$};
        \node(0out)[right of=C]{$\0$};
        
        \draw[arrow] (0in) -- node [midway,above] {0} (A) ;
        \draw[arrow] (A)-- node [midway,left] {1} (B);
        \draw[arrow] (center.center) -- (C) ;
        \draw[arrow] (C) -- node [midway,above] {3} (0out) ;
        \draw[thick] (A) -- (center.center);
        \draw[thick] (B) -- (center.center);
    \end{tikzpicture}
    \caption{Chemical reaction network}
    \end{subfigure}%
    \begin{subfigure}{0.46\textwidth}
    \centering
    \begin{tikzpicture}[node distance=1.2cm]
        \tikzstyle{arrow} = [thick,->,>=stealth]
        \node(A)[]{$A$};
        \node(B)[right of=A]{$B$};
        \node(C)[right of=B]{$C$};
        \node(0)[below of=B]{$\0$};
        
        \draw[arrow] (A).. controls ++(40:9mm) and ++(140:9mm) .. (C);
        \draw[arrow] (A.10) -- (B.170) ;
        \draw[arrow] (B.-170) -- (A.-10) ;
        \draw[arrow] (B) -- (C) ;
        \draw[arrow] (A) -- (0) ;
        \draw[arrow] (B) -- (0) ;
        \draw[arrow] (C) -- (0) ;
    \end{tikzpicture}
    \caption{The influence graph $\G$.}
    \end{subfigure}
    \caption{Example of a simple chemical reaction network and its influence graph. The edges of the influence graph are determined by the non-zero entries of (\ref{eq:example_lap}).}
    \label{fig:example}
\end{figure}
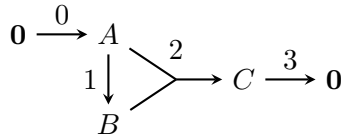
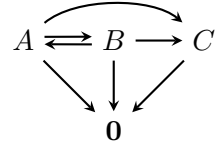
The stoichiometric matrix $S$ and the derivative matrix $R$ are given by
\begin{equation*}
    S = 
    \bordermatrix{
      & 0 & 1 & 2 & 3  \cr
    A & 1 &-1 &-1 & 0  \cr
    B & 0 & 1 &-1 & 0  \cr
    C & 0 & 0 & 1 &-1  \cr
    }, \qquad
    R =
    \bordermatrix{
      & A & B & C  \cr
    0 & 0 & 0 & 0 \cr
    1 & R_{1A} & 0 & 0 \cr
    2 & R_{2A} & R_{2B} & 0  \cr
    3 & 0 & 0 & R_{3C} \cr
    }
\end{equation*}
The generalized Laplacian $\L$ as defined in \ref{eq:L_block} is given by
\begin{equation}\label{eq:example_lap}
\L = \left(\begin{array}{c|c c c}
    0 & R_{2A} & R_{2B} & R_{3C} \\ \hline
    0 & -R_{1A} - R_{2A} & -R_{2B} & 0 \\
    0 & R_{1A} - R_{2A} & -R_{2B} & 0 \\ 
    0 & R_{2A} & R_{2B} & -R_{3C}  
\end{array}\right).
\end{equation}
The influence graph $\G=(\V, \E)$ is the underlying graph of $\L$. Its vertex set and edge set are given by
\begin{align*}
\begin{split}
    \V&=\{A, B, C, \0\}\\
    \E&=\{(A\0), (AB), (AC), (B\0), (BA), (BC), (C\0)\}
\end{split}
\end{align*}
See Figure \ref{fig:example}(b) for a drawing of $\G$. The weights of $\G=(\V,\E, \L)$ are given by the off-diagonal entries of $\L$. Note that the weigth of the edge $(AB)$ has the weight $R_{1A}-R_{2A}$ and hence does not consist of a single variable. Also the variables $R_{jm}$ no longer appear only once, but in multiple edges. This makes analyzing these graphs much harder.

A key difference between the reaction network and its influence graph is that while the reaction network may contain reactions between complexes, e.g.~$A+B\to C$, the influence graph only contains single reactants as nodes and directed edges between them. This is contrary to the approach of writing a chemical reaction network as a directed graph on the set of its complexes which is much more common \cite{Feinberg}. The reaction $j_2:A+B\to C$ results in a total of 6 edges in the influence graph, namely $(A\0),(AC),(B\0),(BC)$ with positive terms and $(AB),(BA)$ with negative terms\footnote{The edge $(AB)$ also gets a positive term from reaction $j_1$.}. This can be interpreted as follows. An increase in the concentration of $A$ accelerates the reaction $j_2$. Hence, $C$ is being produced and we expect its concentration to go up. On the other hand, $B$ is consumed faster and we expect its concentration to decrease. Therefore $(AC)$ gets a positive term while $(AB)$ has a negative term. The same applies for an increase in the concentration of $B$. The edges $(A\0)$ and $(B\0)$ indicate that 'mass is lost' in the reaction $j_2$ meaning that there is only one output while there were two inputs. Since mass conservation is essential for applying generalized Laplacians we add an abstract flow from $A$ and $B$ to $\0$.
\QED\end{example} 
\begin{remark}
    It is convenient to model edges with more than one variable in their weight as multiedges, i.e.~multiple edges with the same start and target node. This turns the influence graph $\G$ into a multigraph. The entire theory of this article can be directly extended to the multigraph setting. However, this is not necessary for the results we present, so for simplicity multigraphs are not considered.
\end{remark}

For a general chemical reaction network, the structure of the influence graph $\G=(\M \cup \{\0\}, \E)$ can be characterized in the following way:
\begin{itemize}
    \item For reactants $m_1\neq m_2$, there is an edge $(m_1m_2)\in \E$ if and only if there is a reaction $j\in \Rea$ in which $m_1$ is an input and $m_2$ is either an input or an output;
    \item For $m \in \M$ there is an edge $(m\0)\in \E$ if and only if $m$ is the input of a reaction $j\in \Rea$ that has an unequal number of inputs and outputs, i.e.~$\1 S^j\neq0$. In particular, this is the case for exit-reaction $m\to \0$.
\end{itemize}
 Note that feed-reactions do not affect this graph in any way as they have no reactant input. We obtain the following result which describes the response of $x^*$ to a reaction perturbation in terms of the influence graph.
\begin{theorem}\label{thm:sensitivity}
    Consider a chemical reaction network that satisfies the non-degeneracy condition $\det(SR)\neq 0$ at equilibrium $x^*$. When perturbing a reaction $j^*\in \Rea$ the response of the equilibrium $x^*$ in a reactant $m'$ is given by
    \begin{equation*}
        \Delta^{j^*}_{m'} = \frac{1}{\norm{\A^\0}} \sum_{\substack{(\T_{m'},\T_\0) \\ \text{dTp}}} \left( \sum_{m\in \T_{m'}} S_{mj^*}\right) \norm{\T_{m'}} \norm{\T_\0}.
    \end{equation*}
    The graph-related terms of the formula refer to the weighted influence graph $\G=(\M \cup \{\0\}, \E, \L)$. We have $\norm{\A^\0}=\det(SR)\neq 0$.
\end{theorem}
\begin{corollary}\label{crl:sensitivity}
    When perturbing a monomolecular reaction $j^*\in \Rea$ of the form $j^*:m^*\to m$ for $m^*, m\in \M \cup \{\0\}$, the response of the equilibrium $x^*$ in a reactant $m'$ is given by
    \begin{equation*}
        \Delta^{j^*}_{m'} = \frac{1}{\norm{\A^\0}} \sum_{\substack{(\T_{m'},\T_\0) \\ j^*\text{-dTp}}} (-1)^{[m^* \in T_{m'}]} \norm{\T_{m'}} \norm{\T_\0},
    \end{equation*}
    where we may interpret $j^*$ as the edge $(m^*m)\in \E$.
\end{corollary}
The proofs to both statements are contained in Section \ref{sec:CRN_proofs}.
\begin{remark}
    Under the assumption $\det(SR)\neq 0$, we find $\norm{\A^\0}\neq 0$. We derive that there is at least one tree rooted in $\0$  in the influence graph and hence there are directed paths in the influence graph from any reactant to $\0$.
\end{remark}
These formulas do not yet provide an efficient way to compute the sensitivity. Instead, they show that the use of the prototypical linear response can be extended to open systems with a more complex network structure. In the special case that each reaction is monomolecular, Corollary \ref{crl:sensitivity} can be used to determine the signs of the sensitivity efficiently. However, this special case has been extensively studied by Vassena \cite{Nico_Sign_Paper} and we will not further elaborate on monomolecular networks. 

\begin{example}
    We revisit Example \ref{ex:CRN} and compute the entries of the response vector $\Delta$ with the help of Theorem \ref{thm:sensitivity}. We start by computing $\norm{\A^\0}$ and, in particular, verifying that $\norm{\A^\0}\neq 0$. In general, the number of trees rooted in $\0$ grows exponentially with the number of nodes. In our case, we can greatly reduce the effort of computing $\norm{\A^\0}$. A tree rooted in $\0$ requires any node to choose one outgoing edge. The node $C$ only has one outgoing edge, while $A$ and $B$ both have $3$. Hence, there are 9 potential combinations to consider, each of which defines a subgraph that might or might not be a tree rooted in $\0$. Such a subgraph is a tree rooted in $\0$ if and only if there are no cycles. In our setting, the only possible cycle is between $A$ and $B$. Therefore, $\norm{\A^\0}$ can be computed by subtracting this particular combination from the sum of all possible combinations, which is a term that factorizes. 
    We compute
    \begin{align*}
        \norm{\A^\0} &= (\ell_{BA} + \ell_{CA} + \ell_{\0A})(\ell_{AB} + \ell_{CB} + \ell_{\0 B}) \ell_{\0 C} - \ell_{BA}\ell_{AB}\ell_{\0C}\\
        &= (R_{1A} - R_{2A} + R_{2A} + R_{2A})(-R_{2B} + R_{2B} + R_{2B})R_{3C} - (R_{1A} - R_{2A})(-R_{2B})R_{3C}\\
        &= 2R_{1A}R_{2B}R_{3C}.
    \end{align*}
    This term is non-zero by assumption.

    We want to compute $\Delta_{m'}^{j^*}$ for all choices of $m'\in \M$ and $j^*\in \Rea$ with the help of Theorem \ref{thm:sensitivity}. It is convenient to fix an $m'\in \M$ and compute $\Delta_{m'}^{j^*}$ for all $j^*\in \Rea$. That way, we only need to determine all divided tree pairs $(\T_{m'}, \T_\0)$ once. \\ \\
    \textbf{Case 1.} $m'=A$
    
    A dTp $(\T_A, T_\0)$ requires us to choose an outgoing edge for the nodes $B$ and $\0$ without forming a cycle. Node $C$ only has one outgoing edge. There is no possibility to create a cycle, hence all three outgoing edges of $B$ result in a dTp.
    \begin{center}
    \begin{tblr}{colspec={Q[t]X[c,m]X[c,m]X[c,m]},
    vlines,}
        \hline
         &
        \SetCell[r=3]{m}{\begin{tikzpicture}[node distance=1cm]
            \tikzstyle{arrow} = [thick,->,>=stealth]
            \node(A)[]{$A$};
            \node(B)[right of=A]{$B$};
            \node(C)[right of=B]{$C$};
            \node(0)[below of=B]{$\0$};
            
            %\draw[arrow] (A).. controls ++(40:9mm) and ++(140:9mm) .. (C);
            \draw[arrow] (B) -- (A) ;
            \draw[arrow] (C) -- (0) ;
        \end{tikzpicture}} &
        \SetCell[r=3]{m}{\begin{tikzpicture}[node distance=1cm]
            \tikzstyle{arrow} = [thick,->,>=stealth]
            \node(A)[]{$A$};
            \node(B)[right of=A]{$B$};
            \node(C)[right of=B]{$C$};
            \node(0)[below of=B]{$\0$};
            
            %\draw[arrow] (A).. controls ++(40:9mm) and ++(140:9mm) .. (C);
            \draw[arrow] (B) -- (C) ;
            \draw[arrow] (C) -- (0) ;
        \end{tikzpicture}} & 
        \SetCell[r=3]{m}{\begin{tikzpicture}[node distance=1cm]
            \tikzstyle{arrow} = [thick,->,>=stealth]
            \node(A)[]{$A$};
            \node(B)[right of=A]{$B$};
            \node(C)[right of=B]{$C$};
            \node(0)[below of=B]{$\0$};
            
            %\draw[arrow] (A).. controls ++(40:9mm) and ++(140:9mm) .. (C);
            \draw[arrow] (B) -- (0) ;
            \draw[arrow] (C) -- (0) ;
        \end{tikzpicture}} \\
        $(\T_A, \T_\0)$ & & & \\
        &&&\\ \hline
        $\norm{\T_A}\norm{\T_\0}$ & $-R_{2B}R_{3C}$ & $R_{2B}R_{3C}$ & $R_{2B}R_{3C}$ \\ \hline
    \end{tblr}
    \end{center}

With this table and Theorem \ref{thm:sensitivity}, we can read off all responses of the form $\Delta_A^{j^*}$ for $j^*\in \Rea$.
\begin{align*}
    \Delta_A^{j_0} &= \frac{1}{2R_{1A}R_{2B}R_{3C}}\big(-R_{2B}R_{3C}+R_{2B}R_{3C} + R_{2B}R_{3C}\big) = \frac{1}{2R_{1A}}\\
    \Delta_A^{j_1} &= \frac{1}{2R_{1A}R_{2B}R_{3C}} \big(0\cdot (-R_{2B}R_{3C})+(-1)\cdot R_{2B}R_{3C} + (-1)\cdot R_{2B}R_{3C}\big) = -\frac{1}{R_{1A}}\\
    \Delta_A^{j_2}&=\frac{1}{2R_{1A}R_{2B}R_{3C}}\big((-2)\cdot(- R_{2B}R_{3C})+(-1)\cdot R_{2B}R_{3C} + (-1)\cdot R_{2B}R_{3C}\big) = 0 \\
    \Delta_A^{j_3} &= \frac{1}{R_{1A}R_{2B}R_{3C}}\big(0\cdot (-R_{2B}R_{3C})+0\cdot R_{2B}R_{3C} + 0\cdot R_{2B}R_{3C}\big) = 0.
\end{align*}
In this case, the signs of the responses are all determined, independent of the precise positive values in $R$.\\ \\
    \textbf{Case 2.} $m'=B$
    
    This case is almost analogous to the previous one.
    \begin{center}
    \begin{tblr}{colspec={Q[t]X[c,m]X[c,m]X[c,m]},
    vlines,}
        \hline
         &
        \SetCell[r=3]{m}{\begin{tikzpicture}[node distance=1cm]
            \tikzstyle{arrow} = [thick,->,>=stealth]
            \node(A)[]{$A$};
            \node(B)[right of=A]{$B$};
            \node(C)[right of=B]{$C$};
            \node(0)[below of=B]{$\0$};
            
            %\draw[arrow] (A).. controls ++(40:9mm) and ++(140:9mm) .. (C);
            \draw[arrow] (A) -- (B) ;
            \draw[arrow] (C) -- (0) ;
        \end{tikzpicture}} &
        \SetCell[r=3]{m}{\begin{tikzpicture}[node distance=1cm]
            \tikzstyle{arrow} = [thick,->,>=stealth]
            \node(A)[]{$A$};
            \node(B)[right of=A]{$B$};
            \node(C)[right of=B]{$C$};
            \node(0)[below of=B]{$\0$};
            
            \draw[arrow] (A).. controls ++(40:9mm) and ++(140:9mm) .. (C);
            \draw[arrow] (C) -- (0) ;
        \end{tikzpicture}} & 
        \SetCell[r=3]{m}{\begin{tikzpicture}[node distance=1cm]
            \tikzstyle{arrow} = [thick,->,>=stealth]
            \node(A)[]{$A$};
            \node(B)[right of=A]{$B$};
            \node(C)[right of=B]{$C$};
            \node(0)[below of=B]{$\0$};
            
            %\draw[arrow] (A).. controls ++(40:9mm) and ++(140:9mm) .. (C);
            \draw[arrow] (A) -- (0) ;
            \draw[arrow] (C) -- (0) ;
        \end{tikzpicture}} \\
        $(\T_B, \T_\0)$ & & & \\
        &&&\\ \hline
        $\norm{\T_B}\norm{\T_\0}$ & $(R_{1A}-R_{2A})R_{3C}$ & $R_{2A}R_{3C}$ & $R_{2A}R_{3C}$ \\ \hline
    \end{tblr}
    \end{center}

With this table and Theorem \ref{thm:sensitivity}, we can read off all responses of the form $\Delta_B^{j^*}$ for $j^*\in \Rea$.
\begin{align*}
    \Delta_B^{j_0} &= \frac{1}{2R_{1A}R_{2B}R_{3C}}\big((R_{1A}-R_{2A})R_{3C} + 0\cdot R_{2A}R_{3C} + 0\cdot R_{2A}R_{3C}\big) = \frac{R_{1A}-R_{2A}}{2R_{1A}R_{2B}}\\
    \Delta_B^{j_1} &= \frac{1}{2R_{1A}R_{2B}R_{3C}} \big(0\cdot (R_{1A}-R_{2A})R_{3C} + R_{2A}R_{3C} + R_{2A}R_{3C}\big) = \frac{R_{2A}}{R_{1A}R_{2B}}\\
    \Delta_B^{j_2}&=\frac{1}{2R_{1A}R_{2B}R_{3C}}\big((-2)\cdot (R_{1A}-R_{2A})R_{3C} + (-1)\cdot R_{2A}R_{3C} + (-1)\cdot R_{2A}R_{3C}\big) = -\frac{1}{R_{2B}} \\
    \Delta_B^{j_3} &= \frac{1}{R_{1A}R_{2B}R_{3C}}\big(0\cdot (R_{1A}-R_{2A})R_{3C} + 0\cdot R_{2A}R_{3C} + 0 \cdot R_{2A}R_{3C}\big) = 0.
\end{align*}
In this case, there is a response entry of undetermined sign. The sign of $\Delta_B^{j_0}$ depends on the inequality $R_{1A} \gtrless R_{2A}$. If $R_{1A}=R_{2A}$, the response $\Delta_B^{j_0}$ is $0$.\\ \\
\textbf{Case 3.} $m'=C$

This case is significantly more tedious than the previous two. For $m'=C$ there are a total of 8 divided tree pairs $(\T_{m'}, \T_\0)$.\\
    \begin{center}
    \begin{tblr}{colspec={Q[t]X[c,m]X[c,m]X[c,m]X[c,m]},
    vlines,}
        \hline
         &
        \SetCell[r=3]{m}{\begin{tikzpicture}[node distance=1cm]
            \tikzstyle{arrow} = [thick,->,>=stealth]
            \node(A)[]{$A$};
            \node(B)[right of=A]{$B$};
            \node(C)[right of=B]{$C$};
            \node(0)[below of=B]{$\0$};
            
            %\draw[arrow] (A).. controls ++(40:9mm) and ++(140:9mm) .. (C);
            \draw[arrow] (A) -- (0) ;
            \draw[arrow] (B) -- (0) ;
        \end{tikzpicture}} &
        \SetCell[r=3]{m}{\begin{tikzpicture}[node distance=1cm]
            \tikzstyle{arrow} = [thick,->,>=stealth]
            \node(A)[]{$A$};
            \node(B)[right of=A]{$B$};
            \node(C)[right of=B]{$C$};
            \node(0)[below of=B]{$\0$};
            
            %\draw[arrow] (A).. controls ++(40:9mm) and ++(140:9mm) .. (C);
            \draw[arrow] (A) -- (0) ;
            \draw[arrow] (B) -- (C) ;
        \end{tikzpicture}} & 
        \SetCell[r=3]{m}{\begin{tikzpicture}[node distance=1cm]
            \tikzstyle{arrow} = [thick,->,>=stealth]
            \node(A)[]{$A$};
            \node(B)[right of=A]{$B$};
            \node(C)[right of=B]{$C$};
            \node(0)[below of=B]{$\0$};
            
            %\draw[arrow] (A).. controls ++(40:9mm) and ++(140:9mm) .. (C);
            \draw[arrow] (A) -- (0) ;
            \draw[arrow] (B) -- (A) ;
        \end{tikzpicture}} & 
        \SetCell[r=3]{m}{\begin{tikzpicture}[node distance=1cm]
            \tikzstyle{arrow} = [thick,->,>=stealth]
            \node(A)[]{$A$};
            \node(B)[right of=A]{$B$};
            \node(C)[right of=B]{$C$};
            \node(0)[below of=B]{$\0$};
            
            %\draw[arrow] (A).. controls ++(40:9mm) and ++(140:9mm) .. (C);
            \draw[arrow] (A) -- (B) ;
            \draw[arrow] (B) -- (0) ;
        \end{tikzpicture}} \\
        $(\T_C, \T_\0)$ & & & &\\
        &&&&\\ \hline
        $\norm{\T_C}\norm{\T_\0}$ & $R_{2A}R_{2B}$ & $R_{2A}R_{2B}$ & $-R_{2A}R_{2B}$ & $(R_{1A} - R_{2A})R_{2B}$\\ \hline \hline
         &
        \SetCell[r=3]{m}{\begin{tikzpicture}[node distance=1cm]
            \tikzstyle{arrow} = [thick,->,>=stealth]
            \node(A)[]{$A$};
            \node(B)[right of=A]{$B$};
            \node(C)[right of=B]{$C$};
            \node(0)[below of=B]{$\0$};
            
            %\draw[arrow] (A).. controls ++(40:9mm) and ++(140:9mm) .. (C);
            \draw[arrow] (A) -- (B) ;
            \draw[arrow] (B) -- (C) ;
        \end{tikzpicture}} &
        \SetCell[r=3]{m}{\begin{tikzpicture}[node distance=1cm]
            \tikzstyle{arrow} = [thick,->,>=stealth]
            \node(A)[]{$A$};
            \node(B)[right of=A]{$B$};
            \node(C)[right of=B]{$C$};
            \node(0)[below of=B]{$\0$};
            
            \draw[arrow] (A).. controls ++(40:9mm) and ++(140:9mm) .. (C);
            %\draw[arrow] (A) -- (0) ;
            \draw[arrow] (B) -- (0) ;
        \end{tikzpicture}} & 
        \SetCell[r=3]{m}{\begin{tikzpicture}[node distance=1cm]
            \tikzstyle{arrow} = [thick,->,>=stealth]
            \node(A)[]{$A$};
            \node(B)[right of=A]{$B$};
            \node(C)[right of=B]{$C$};
            \node(0)[below of=B]{$\0$};
            
            \draw[arrow] (A).. controls ++(40:9mm) and ++(140:9mm) .. (C);
            %\draw[arrow] (A) -- (0) ;
            \draw[arrow] (B) -- (C) ;
        \end{tikzpicture}} & 
        \SetCell[r=3]{m}{\begin{tikzpicture}[node distance=1cm]
            \tikzstyle{arrow} = [thick,->,>=stealth]
            \node(A)[]{$A$};
            \node(B)[right of=A]{$B$};
            \node(C)[right of=B]{$C$};
            \node(0)[below of=B]{$\0$};
            
            \draw[arrow] (A).. controls ++(40:9mm) and ++(140:9mm) .. (C);
            %\draw[arrow] (B) -- (A) ;
            \draw[arrow] (B) -- (A) ;
        \end{tikzpicture}} \\
        $(\T_C, \T_\0)$ & & & &\\
        &&&&\\ \hline
        $\norm{\T_C}\norm{\T_\0}$ & $(R_{1A} - R_{2A})R_{2B}$ & $R_{2A}R_{2B}$ & $R_{2A}R_{2B}$ & $-R_{2A}R_{2B}$\\ \hline
    \end{tblr}
    \end{center}
    
    We will not write out entire calculations of the response entries as before. We compute
    \begin{align*}
        \Delta_C^{j_0} &= \frac{1}{2R_{3C}}
        &\Delta_C^{j_1} &= 0\\
        \Delta_C^{j_2} &= 0 
        &\Delta_C^{j_3} &= -\frac{1}{R_{3C}}
    \end{align*}
    
    These 4 responses have an intuitive explanation. For the chemical reaction network we consider it makes sense to associate a 'mass' of $1$ to both reactants $A$ and $B$ and a mass of $2$ to reactant $C$. For this assignment of mass, the inner reaction $j_1$ and $j_2$ are mass preserving. We deduce that the total inflow and outflow of mass of the reaction network have to be equal. In particular, the reaction rates of $j_0$ and $j_3$ must satisfy \begin{equation}\label{eq:in_out}
        r_{j_0} = 2 r_{j_3}(x^*_C),
    \end{equation}
    where $r_{j_0}$ is the constant rate of the inflow reaction $j_0$ and $r_{j_3}(x^*_C)$ is the rate of the outflow reaction $j_3$ at equilibrium $x^*$.
    Since $C$ is the only input of $j_3$ the concentration of $C$ directly determines the rate $r_{j_3}$. Hence, the concentration of $C$ cannot change as long as the inflow stays the same. There are only two ways to achieve a response in the concentration of $C$. Either by adding additional inflow via reaction $j_0$ or by accelerating the outflow reaction $j_3$ and thereby requiring a lower concentration of $C$ to obtain the same outflow. This explains why $\Delta_C^{j_1}$ and $\Delta_C^{j_2}$ are 0. The terms for $\Delta_C^{j_1}$ and $\Delta_C^{j_2}$ can be derived by adding and $\epsilon$ to either one side of (\ref{eq:in_out}).
\end{example}\QED

\subsection{Proofs}\label{sec:CRN_proofs}
\textit{Proof of Theorem \ref{thm:sensitivity}:} To prove this theorem, we build a connection between the sensitivity of reaction networks and the prototypical linear response of generalized Laplacians. First, we find that $\L$ has at least one solution to the equilibrium equation (\ref{eq:proto_equation}), which is given by $\mu=e_\0$ (compare (\ref{eq:L_block})). One can check that $\det\big( \Li{\0} \big)=\det(SR)\neq 0$ such that $\mu$ is in fact the unique solution to the equilibrium equation by Proposition \ref{prop:det_Li}. We show that the response $\delta$ of $\mu$ to certain perturbations in $\L$ has the same form as the responses $\Delta$ of the reaction network. We start with the case of perturbing an entry of the form $\L_{m\0}$ and studying the response vector $\delta^{j_m}\in \R^{\M \cup \{\0\}}$ as defined in (\ref{eq:response_general}). We obtain from (\ref{eq:implicit_diff}), the implicit differentiation of $\mu(\epsilon)$,
\begin{equation*}
    \delta^{j_m} =- \left( \Li{\0} \right)^{-1} (\mu_\0 e_{m}).
\end{equation*}
Since $\mu_\0=1$, the factor can be omitted. The inverse of $\Li{\0}$ is easily verified to be given by
\begin{equation*}
\left( \Li{\0} \right)^{-1} = \left(\begin{array}{c|c c c}
    1 & \multicolumn{3}{c}{-\1 (SR)^{-1}} \\ \hline
    \multirow{3}{*}{0} & & & \\
    & & (SR)^{-1} & \\ 
    & & & 
\end{array}\right).
\end{equation*}
We find that the response vector $\delta^{j_m}$ restricted to the set of reactants $\M$ has a similar form as the response vectors in the reaction network (compare (\ref{eq:metabolite_response}))
\begin{equation*}
    \delta^{j_m}|_\M =- (SR)^{-1} e_{m}.
\end{equation*}

\iffalse
This already lets us describe the response to a feed-reaction perturbation with the help of Theorem \ref{thm:general_main}. We compute the response in $m'\in \M$ to a perturbation of a feed-reaction $j^*:\0 \to m$
\begin{align*}
    \Delta^{j^*}_{m'} = \mu_{\0} \frac{1}{\norm{\A}} \sum_{w\neq m'} \sum_{\substack{(\T_{m'},\T_w) \\ j^*\text{-dTp}}} (-1)^{[\0 \in \T_{m'}]} \norm{\T_{m'}}\norm{\T_w}.
\end{align*}
Remember that $\mu=e_\0$ and therefore $\mu_\0 = 1$. In the influence graph the node $\0$ has no outgoing edges, so $\norm{\A}=\norm{\A^\0}$ since any rooted spanning tree has to be rooted in $\0$. For the same reason in any $j^*$-dTp one of the trees must be rooted in $\0$. Hence, there only exist $j^*$-dTp for $w=\0$. We obtain 
\begin{equation*}
    \Delta^{j^*}_{m'} = \frac{1}{\norm{\A^\0}} \sum_{\substack{(\T_{m'},\T_\0) \\ j^*\text{-dTp}}} (-1)^{[\0 \in \T_{m'}]} \norm{\T_{m'}}\norm{\T_\0}
\end{equation*}
Lastly, we notice that $\0$ can never be in $\T_{m'}$ as it is the second root and the $(-1)$ factor vanishes. Additionally, $(\T_{m'}, \T_\0)$ being a $j^*$-dTp is equivalent to $(\T_{m'}, \T_\0)$ being a dTp with $m\in \T_{m'}$ (compare Definition \ref{def:j-dTp}). We conclude
\begin{equation}\label{eq:feed_response}
    \Delta^{j^*}_{m'}
    =\frac{1}{\norm{\A^\0}} \sum_{\substack{(\T_{m'},\T_\0) \\ \text{dTp}}} [m\in \T_{m'}] \norm{\T_{m'}}\norm{\T_\0}.
\end{equation}
This formula describes the response in $m'\in \M$ to the perturbation of a feed-reaction $j^*:\0\to m$. It will be the basis for any further sensitivity calculations.\fi

We observe from (\ref{eq:metabolite_response}) that the response to the perturbation of a reaction $j^*\in \Rea$ can be decomposed into summands of this form
\begin{align*}
    \Delta^{j^*}=-(SR)^{-1}S^{j^*} &= \sum_{m\in \M} S_{mj^*} \left( -(SR)^{-1} e_m\right).
\end{align*}
Using the formula of Theorem \ref{thm:general_main} to describe the entries of $\delta^{j_m}$, we compute for $m'\in \M$
\begin{align*}
    \Delta^{j^*}_{m'}
    &= \sum_{m\in \M} S_{mj^*} \left( -(SR)^{-1} e_m\right)_{m'} \\
    &= \sum_{m\in \M} S_{mj^*} \delta^{j_m}_{m'} \\
    &= \sum_{m\in \M} S_{mj^*} \left(\frac{1}{\norm{\A^\0}} \sum_{\substack{(\T_{m'},\T_\0) \\ \text{dTp}}} [m\in \T_{m'}] \norm{\T_{m'}}\norm{\T_\0}\right)\\
    &= \frac{1}{\norm{\A^\0}} \sum_{\substack{(\T_{m'},\T_\0) \\ \text{dTp}}} \left(\sum_{m\in \M} S_{mj^*}[m\in \T_{m'}] \right)\norm{\T_{m'}}\norm{\T_\0} \\
    &=\frac{1}{\norm{\A^\0}} \sum_{\substack{(\T_{m'},\T_\0) \\ \text{dTp}}} \left(\sum_{m\in \T_{m'}} S_{mj^*}\right)\norm{\T_{m'}}\norm{\T_\0}.
\end{align*}
This completes the proof of Theorem \ref{thm:sensitivity}.\QED

\textit{Proof of Corollary \ref{crl:sensitivity}:} To prove the Corollary, consider a monomolecular reaction $j^*:m^*\to m\in \Rea$. Theorem \ref{thm:sensitivity} provides
\begin{equation*}
    \Delta^{j^*}_{m'}
    = \frac{1}{\norm{\A^\0}} \sum_{\substack{(\T_{m'},\T_\0) \\ \text{dTp}}} \left(\sum_{m\in \T_{m'}} S_{mj^*}\right)\norm{\T_{m'}}\norm{\T_\0}.
\end{equation*}
The vector $S^{j^*}=e_{m} - e_{m^*}$ only has two non-zero-entries and therefore
\begin{equation*}
    \Delta^{j^*}_{m'}
    = \frac{1}{\norm{\A^\0}} \sum_{\substack{(\T_{m'},\T_\0) \\ \text{dTp}}} \big([m\in \T_{m'}] - [m^*\in \T_{m'}] \big)\norm{\T_{m'}}\norm{\T_\0}.
\end{equation*}
The middle part is only non-zero if exactly one of the nodes $m^*, m$ is in $\T_{m'}$. By Definition \ref{def:j-dTp} this is equivalent to $(\T_{m'},\T_\0)$ being a $j^*$-dTp. The factor is $-1$ if $m^*\in \T_{m'}$ and $+1$ if $m^*\notin \T_{m'}$. We conclude
\begin{equation*}
        \Delta^{j^*}_{m'} = \frac{1}{\norm{\A^\0}} \sum_{\substack{(\T_{m'},\T_\0) \\ j^*\text{-dTp}}} (-1)^{[m^* \in T_{m'}]} \norm{\T_{m'}} \norm{\T_\0}.
\end{equation*}
This concludes the proof.\\
\QED

\section{Outlook}\label{sec:outlook}
Analyzing the sensitivity of an equilibrium is one of the key steps in approaching the control problem, i.e.~achieving a desired response equilibrium by perturbing particular system parameters. In the case of CTMCs on finite state spaces that admit a unique stat.~dist.~Theorem \ref{thm:main_thm} provides a solid basis for this setting. However, to achieve a desired response in the stat.~dist.~it is in general necessary to perturb more than one transition rate. This may result in cancellation when the sign of two responses differ. Understanding these cancellations is crucial in order to effectively apply the results we presented to the control problem.

The final goal of the sensitivity study of chemical reaction networks as we carried it out is to precisely link the algebraic sign of the response to the underlying network structure, similar to Corollary \ref{crl:signed_sens}. The formula of Theorem \ref{thm:sensitivity} reduced this problem to determining whether positive/negative summands occur. However, determining the sign of a summand is non-trivial for general networks. The difficulty does not necessarily lie in the existence of both positive and negative rates, but rather in the fact that the same variable may occur multiple times in the generalized Laplacian (compare (\ref{eq:example_lap})). For monomolecular networks, this is not an issue and the algebraic sign of the sensitivity can be characterized precisely from the network structure. Such a result for monomolecular networks can be found in \cite{Nico_Sign_Paper}. It is not ruled out that for general reaction networks determining the algebraic signs of the sensitivity based on the graph structure is in NP (assuming P$\neq$NP) such that a simple characterization is impossible. In that case, further assumptions on the network, e.g. ruling out catalytic reactions (reaction in which a reactant is both input and output), could make a 'simple' characterization possible. In order to find these suitable assumptions, one needs to study which conditions the influence graph of a reaction network needs to satisfy such that determining the signs in Theorem \ref{thm:sensitivity} becomes easier.

\section{Acknowledgements}
I thank Bernold Fiedler and Nicola Vassena for introducing me to the presented topic as well as motivating many of the results and methods used. Special thanks go to Maximilian Engel and Dennis Chemnitz for their support by reading through the manuscript and refining its structure. My work on this topic has been funded by the Berlin Mathematical School.
\bibliographystyle{unsrt}
\bibliography{bibliography}

\begin{thebibliography}{10}

\bibitem{barzel2013universality}
Baruch Barzel and Albert-L{\'a}szl{\'o} Barab{\'a}si.
\newblock Universality in network dynamics.
\newblock {\em Nature physics}, 9(10):673--681, 2013.

\bibitem{nakajima1992sensitivity}
Hisao Nakajima.
\newblock Sensitivity and stability of flow networks.
\newblock {\em Ecological Modelling}, 62(1-3):123--133, 1992.

\bibitem{yodzis1988indeterminacy}
Peter Yodzis.
\newblock The indeterminacy of ecological interactions as perceived through
  perturbation experiments.
\newblock {\em Ecology}, 69(2):508--515, 1988.

\bibitem{Feinberg}
Martin Feinberg.
\newblock Chemical reaction network structure and the stability of complex
  isothermal reactors—i. the deficiency zero and deficiency one theorems.
\newblock {\em Chemical Engineering Science}, 42:2229--2268, 1987.

\bibitem{Brehm_Fiedler3}
Bernhard Brehm and Bernold Fiedler.
\newblock Sensitivity of chemical reaction networks: A structural approach.
  3. regular multimolecular systems.
\newblock {\em Mathematical Methods in the Applied Sciences}, 41(4):1344--1376,
  2018.

\bibitem{Nico_thesis}
Nicola Vassena.
\newblock Sensitivity of metabolic networks.
\newblock {\em PhD Thesis, Freie Universität Berlin}, 2020.

\bibitem{feliu2019sign}
Elisenda Feliu.
\newblock Sign-sensitivities for reaction networks: an algebraic approach.
\newblock {\em Mathematical Biosciences and Engineering}, 16(6):8195--8213,
  2019.

\bibitem{zahmati2010steady}
Amir~Sepasi Zahmati, Xavier Fernando, and Ali Grami.
\newblock Steady-state markov chain analysis for heterogeneous cognitive radio
  networks.
\newblock In {\em 2010 IEEE Sarnoff Symposium}, pages 1--5. IEEE, 2010.

\bibitem{schweitzer_discrete_sens}
Paul~J. Schweitzer.
\newblock Perturbation theory and finite {M}arkov chains.
\newblock {\em Journal of Applied Probability}, 5(2):401--413, 1968.

\bibitem{hunter1986stationary}
Jeffrey~J Hunter.
\newblock Stationary distributions of perturbed markov chains.
\newblock {\em Linear Algebra and its applications}, 82:201--214, 1986.

\bibitem{funderlic1986sensitivity}
Robert~E Funderlic and CD~Meyer~Jr.
\newblock Sensitivity of the stationary distribution vector for an ergodic
  markov chain.
\newblock {\em Linear Algebra and its Applications}, 76:1--17, 1986.

\bibitem{wang2019steady}
Ting Wang and Petr Plechac.
\newblock Steady-state sensitivity analysis of continuous time markov chains.
\newblock {\em SIAM Journal on Numerical Analysis}, 57(1):192--217, 2019.

\bibitem{Mochizuki_Fiedler1}
Atsushi Mochizuki and Bernold Fiedler.
\newblock Sensitivity of chemical reaction networks: A structural approach. 1.
  examples and the carbon metabolic network.
\newblock {\em Journal of theoretical biology}, 367, 11 2014.

\bibitem{fournier_graphs}
Jean-Claude Fournier.
\newblock {\em Graphs theory and applications: with exercises and problems}.
\newblock John Wiley \& Sons, 2013.

\bibitem{MCTT}
V.~Anantharam and P.~Tsoucas.
\newblock A proof of the markov chain tree theorem.
\newblock {\em Statistics \& Probability Letters}, 8(2):189--192, 1989.

\bibitem{2_trees}
Wai-Kai Chen.
\newblock {\em Applied graph theory}, volume~13.
\newblock Elsevier, 2012.

\bibitem{Chaiken_AMMTT}
Seth Chaiken.
\newblock A combinatorial proof of the all minors matrix tree theorem.
\newblock {\em SIAM Journal on Algebraic Discrete Methods}, 3(3):319--329,
  1982.

\bibitem{Stroock2014CTMC}
Daniel~W. Stroock.
\newblock {\em Markov Processes in Continuous Time}, pages 99--136.
\newblock Springer Berlin Heidelberg, Berlin, Heidelberg, 2014.

\bibitem{norris1998markov}
James~R Norris.
\newblock {\em Markov chains}.
\newblock Number~2. Cambridge university press, 1998.

\bibitem{Fiedler_Mochizuki2}
Bernold Fiedler and Atsushi Mochizuki.
\newblock Sensitivity of chemical reaction networks: a structural approach. 2.
  regular monomolecular systems.
\newblock {\em Mathematical Methods in the Applied Sciences}, 38, 05 2015.

\bibitem{Nico_Sign_Paper}
Nicola Vassena.
\newblock Sensitivity of monomolecular reaction networks: Signed flux-response
  to reaction rate perturbations.
\newblock In {\em 2017 European Conference on Circuit Theory and Design
  (ECCTD)}, pages 1--4, 2017.

\end{thebibliography}

\end{document}